\documentclass[10pt,twocolumn,floatfix,aps,prl,nofootinbib,superscriptaddress]{revtex4-2}

\usepackage{amsmath, amssymb, graphicx, hyperref, xcolor, colortbl, braket}
\usepackage{cleveref}
\usepackage{comment}
\usepackage{mathtools}
\usepackage{titlesec}
\usepackage{tikz}
\usepackage{tkz-euclide}

\usetikzlibrary{arrows.meta,decorations.markings}

\DeclareMathOperator{\Sym}{Sym}
\DeclareMathOperator{\Span}{span}
\DeclareMathOperator{\Tr}{tr}

\newcommand\nx{{n_{\!X}}}
\newcommand\ny{{n_{\!Y}}}
\newcommand\nz{{n_{\!Z}}}

\DeclareMathOperator*{\mediumotimes}{\text{\raisebox{0.25ex}{\scalebox{0.8}{$\bigotimes$}}}}

\crefname{equation}{eq.}{eqs.}
\Crefname{equation}{Eq.}{Eqs.}
\crefname{figure}{fig.}{figs.}
\Crefname{figure}{Fig.}{Figs.}

\begin{document}

\title{Symmetric tensor scars with tunable entanglement from volume to area law}

\author{Bhaskar Mukherjee}
\affiliation{S. N. Bose National Centre for Basic Sciences, Block JD, Sector III, Salt Lake, Kolkata 700106, India}

\affiliation{Department of Physics and Astronomy, University College London, Gower Street, London, WC1E 6BT, UK}

\author{Christopher J. Turner}
\affiliation{Department of Physics and Astronomy, University College London, Gower Street, London, WC1E 6BT, UK}

\author{Marcin Szyniszewski}
\affiliation{Department of Physics and Astronomy, University College London, Gower Street, London, WC1E 6BT, UK}
\affiliation{Department of Computer Science, University of Oxford, Parks Road, Oxford OX1 3QD, UK}

\author{Arijeet Pal}
\affiliation{Department of Physics and Astronomy, University College London, Gower Street, London, WC1E 6BT, UK}

\begin{abstract}
Teleportation of quantum information over long distances requires robust entanglement on the macroscopic scale. The construction of highly energetic eigenstates with tunable long-range entanglement can provide a new medium for information transmission. Using a symmetric superposition of the antipodal triplet states, we construct polynomially many exact zero-energy eigenstates for a class of non-integrable spin-1/2 Hamiltonians with two-body interactions. These states exhibit non-thermal correlations, hence, are genuine quantum many-body scars. By tuning the distribution of triplets we induce extensive, logarithmic, or area-law entanglement, and can observe a second-order entanglement phase transition. Quasiparticle excitations in this manifold converge to be exact quantum many-body scars in the thermodynamic limit. This framework has a natural extension to higher dimensions, where entangled states controlled by lattice geometry and internal symmetries can result in new classes of correlated out-of-equilibrium quantum matter. Our results provide a new avenue for entanglement control and quantum state constructions. 
\end{abstract}

{\maketitle}

\titleformat{\section}[runin]{\normalfont\itshape}{\thesection}{0ex}{\phantomsection}[.---]
\titlespacing{\section}{\parindent}{0ex}{0ex}

\section{Introduction}
It was previously conjectured that all thermalizing non-integrable quantum many-body systems satisfy a strong eigenstate thermalization hypothesis (ETH)~\cite{Rigol2008, DAlessio2016, strongETH}, wherein initial conditions generically thermalize and are unable to retain long-range quantum correlations.
In recent years, many exceptions have been found in a wide range of models, featuring ETH-violating eigenstates, known as quantum many-body scars (QMBS)~\cite{Turner:2018a, Turner:2018b, Moudgalya_2018_2, Lin_2019, Iadecola2020, Lee2020, Wildeboer2021, Srivatsa2020, serbyn2021QMBS, Richter2022, sanjay_review, quasiparticle_scar_review}, following a ground-breaking experiment in a Rydberg atom quantum simulator~\cite{Bernien2017}.
This phenomenon can dramatically slow down thermalization for weakly entangled initial states and preserve quantum correlations, including topological order~\cite{Wildeboer2021,Srivatsa2020,Jeyaretnam2021Jul}.

Traditionally, these examples have featured area-law or logarithmic entanglement scaling, which can be understood as an embedding of ground states of a particular Hamiltonian deep into the many-body spectrum of a non-integrable model~\cite{Shiraishi:2017}.
More recently, eigenstates with volume-law entanglement have attracted significant interest as they usually satisfy ETH. Nonetheless, atypical volume-law states can encode quantum information in accessible operators which can be robust despite being highly entangled. Understanding the exact structure of these states can help identify observables that behave athermally hidden in the complex entanglement features, and can potentially be utilized for information storage and communication. In this respect, Bell pairs have proven to be useful building blocks for such highly entangled states. Rainbow scars~\cite{rainbow_scars,rainbow1,rainbow2,rainbow3} are a particular example formed by a concentric arrangement of Bell pairs over arbitrary distances, which exhibit a volume law for a large majority of bipartitions.
Similar constructions play a role in preparing thermofield double states~\cite{Juan_Maldacena_2003, Hartmann, Raju, Cottrell} describing the interior of black holes. In periodic spin chains, volume-law states can also be prepared using
entangled antipodal pairs of spins~\cite{EAP, Mohapatra2024Oct, Ivanov2024Mar} both in integrable~\cite{Crosscap1, Gombor2022Jul, crosscap2} and non-integrable models~\cite{Udupa2023Dec}. Such states are also known as ``crosscap" states in conformal field theory for a long time~\cite{Ishibashi}.

Many-body states formed by singlet coverings have a rich history in condensed matter physics. The celebrated Majumdar-Ghosh state~\cite{MG1, MG2} is the frustration-free ground state of
a Heisenberg-type spin chain.
This state consists of a nearest-neighbor (n.n.) singlet covering and exhibits valence bond order. In 2D such order can be destroyed by superposing all n.n.\@ singlet coverings, forming a quantum spin liquid~\cite{Anderson1973Feb, Anderson1987Mar, Baskaran1987, Baskaran1993Dec}, realized at the Rokhsar-Kivelson point in quantum dimer models~\cite{RK, RKS, RoderichSondhi}.
The singlets can be long-ranged as well~\cite{Balents2010Mar}. In fact, a spin liquid can be obtained from an ordered state by tuning the length distribution of the singlets~\cite{tuneRVB}. The possibility of realizing states that are a superposition of Bell pair coverings at higher energies, particularly at infinite temperature, has attracted limited attention. 

Quantum many-body systems with spectral reflection symmetry can host a zero-energy manifold, exponentially large in system size~\cite{Turner:2018a,Turner:2018b,Schecter_Chiral_Index_2018}.
These states lie at infinite temperature and are expected to satisfy ETH.
Nevertheless, a highly degenerate manifold offers the possibility of constructing QMBS by a suitable superposition of states in this manifold~\cite{Lin_2019,Karle_AreaLaw, senZeromodes}.
Staggered Hamiltonians have also recently emerged as possible hosts of quantum scars~\cite{Lee2020, Turner2024Jul, Swain2021, Melendrez2025Jan}.
In this letter, we construct a large variety of exact zero-energy eigenstates for a class of staggered Heisenberg models in one dimension by superposing long-range triplet coverings. Their entanglement can be tuned from volume-law to area-law by controlling the distribution of triplets. Our construction is generalizable to higher dimensions and is stable to several forms of symmetry-breaking perturbations. 

\section{Model and the symmetric tensor scars}
Consider a class of bond-staggered Heisenberg Hamiltonians with long-range interactions on a periodic chain of size $2N$,
\begin{equation}
    H = \sum_{i=1}^{2N} (-1)^i {\bf S}_i \cdot {\bf S}_{i+r},
    \label{eq:SH}
\end{equation}
where ${\bf S} = (\sigma_X,\sigma_Y,\sigma_Z) / 2$ is a vector of spin-1/2 operators, and $r\in[0,N-1]$ is the range of interactions.
$H$ is invariant under translation by two lattice sites, reflection over a bond, and has an internal SU(2) symmetry. The nearest-neighbor version of the model has a long history, the ground state belongs to the Haldane phase of gapped symmetry protected topological order \cite{hida_1992,hida_1992_phasediagram,Kennedy1992} which was also realized experimentally \cite{manaka1997}.

\begin{figure}
  \includegraphics{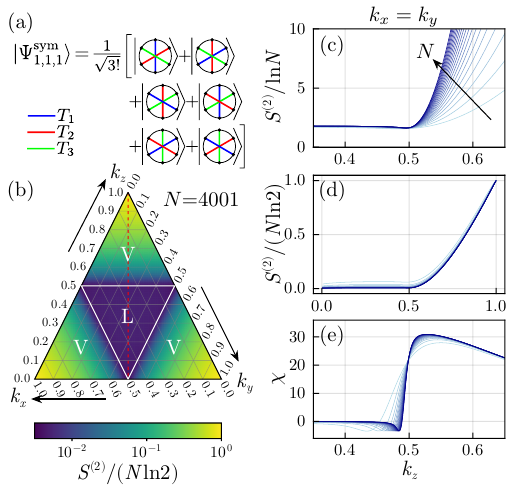}
  \caption{
    (a)~A symmetric tensor state for $N=3$, a zero-energy eigenstate of \Cref{eq:SH}. Each component is a tensor product of three triplet types ($T_1$, $T_2$, $T_3$) between antipodal sites.
    (b)~Phase diagram for the Bell basis states. Colour indicates the half-system entropy $S^{(2)}$ for $N{=}4001$. White lines indicate the boundary between volume-scaling entanglement (V) and log-scaling entanglement phases (L).
    The red dashed line indicates $k_x {=} k_y$ which is shown in more detail in panels (c-e), wherein increasing saturation indicates increasing $N$ from 201 to 4001, in steps of 200.
    Along this line, we show entanglement rescaled by (c)~$\ln N$ and (d)~$N$ to demonstrate the two scalings.
    (e)~Susceptability $\chi = (1/N)\mathrm{d}^2 S^{(2)} / \mathrm{d} k_z^2$ exhibits a crossing evidencing a second-order phase transition.
  }
  \label{fig:Bell}
\end{figure}

We define the following many-body states, which we designate as \emph{root states},
\begin{equation}
    \ket{\Psi(v)} = v^{\otimes N},
    \label{eq:singlet}
\end{equation}
where $v$ is a state of two spins at antipodal sites $i$ and $i+N$. 
We find that \Cref{eq:singlet} is a zero-energy eigenstate of \Cref{eq:SH} provided that $N$ is odd and $v$ is either the singlet state or any triplet state~\cite{SuppMat}.
The singlet root state $\ket{\Psi(S)}$ belongs to $S^2_\text{tot} = 0$, the largest symmetry sector of \Cref{eq:SH} with maximal magnon number, where $N_{\text{mag}}=N-\langle |S^z_{\text{tot}}|\rangle$ (deficit in total spin polarization captures the average number of magnons in the state).
Construction of exact zero energy eigenstates of \Cref{eq:SH} with high magnon number has been a recent challenge~\cite{Turner2024Jul,Melendrez2025Jan} which we overcome in this letter.
The root states $\ket{\Psi(T)}$, constructed from some triplet representation $T$, are generically not eigenstates of $S^z_{\text{tot}}$ and have no well-defined magnon numbers, but will typically have an extensive average number of magnons. We note that certain root states can also be annihilated by infinitely many odd conserved charges of the XXX spin chain~\cite{Crosscap1}; hence, they can be treated as integrable boundary states to construct new scar models~\cite{IBS1, IBS2}.

The linear span of the root states (excluding the singlet root state)
is isomorphic to the space of symmetric tensors of rank-$N$ over the 3-dimensional vector space of triplets, $\text{Sym}^{N}(V_{S=1})$, with dimension $\mathcal{N} = \binom{N{+}2}{N} = (N^2 + 3N + 2)/2$.
Hence, this subspace scales quadratically with the system size and is much smaller than the exponentially large space of all zero-energy states.

To construct a family of complex zero-energy eigenstates, we take the innovative step of introducing superpositions of triplet coverings.
We show that choosing different triplets for the antipodal pairs and organizing them in a symmetric linear superposition, results in a family of exact zero-energy eigenstates with an intricate web of quantum correlations.
These states form a complete basis for the \emph{symmetric tensor scars} and
can be labeled by three positive integers, $n_1, n_2, n_3$ (the number of triplets of each kind), which sum to $N$, and are formally represented as,
\begin{align}
    \ket{\Psi^\text{sym}_{n_1,n_2,n_3}}
    &{=} \frac{1}{{\mathcal{N}_c}}
    \smashoperator{\sum_{\pi \in S_{N}}}
    \pi (\ket{T_1}^{\otimes n_1} {\otimes} \ket{T_2}^{\otimes n_2} {\otimes} \ket{T_3}^{\otimes n_3})
    \text{,}
    \label{eq:soln2}
\end{align}
where $\pi(\cdot)$ is a permutation of the tensor-power factors (an element of the symmetric group of $N$ objects, $S_N$).
The normalization $\mathcal{N}_c = \sqrt{N!\,n_1!\,n_2!\,n_3!}$ can be computed from the orbit-stabilizer theorem.
We show one such state in \Cref{fig:Bell}(a).
The anomalous nature of these states (scars) will become evident shortly.

Although there are infinitely many choices for the triplet basis, here we consider the following two, which serve as examples of the most and the least entangled states,
\begin{subequations}
\begin{align}
    \ket{T^B_X}_i &= \tfrac{1}{\sqrt{2}} (\ket{\uparrow_i\uparrow_j} + \ket{\downarrow_i\downarrow_j}) \ \ ; \ \ \ket{T^C_+}_i=\ket{\uparrow_i\uparrow_j},\\
    \ket{T^B_Y}_i &= \tfrac{1}{\sqrt{2}}(\ket{\uparrow_i\uparrow_j} - \ket{\downarrow_i\downarrow_j})  \ \ ; \ \ \ket{T^C_-}_i=\ket{\downarrow_i\downarrow_j},\\
    \ket{T^B_Z}_i &= \tfrac{1}{\sqrt{2}} (\ket{\uparrow_i\downarrow_j} + \ket{\downarrow_i\uparrow_j}) \ \ ; \ \ \ket{T^C_Z}_i=\ket{T^B_Z}_i,
\end{align}
\end{subequations}
where $j\equiv i+N$, while $\ket{\uparrow}$ and $\ket{\downarrow}$ are eigenvectors of $S^z$.
We call $T^B$ ($T^C$) the Bell pair (conventional) basis for the triplets. Subsequently, we represent a family of symmetric tensor states in these two bases by $(n_X,n_Y,n_Z)$ and $(n_+,n_-,n_Z)$, respectively.

\section{Entanglement entropy}
\begin{figure}
\includegraphics[width=\linewidth]{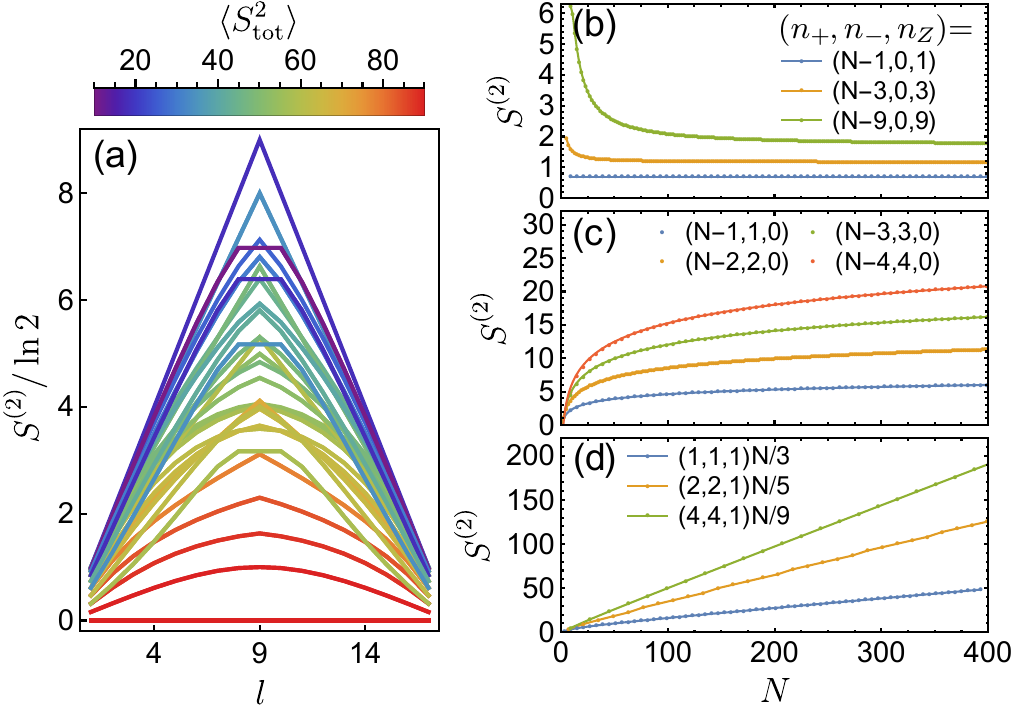}
\caption{Entanglement for the symmetric tensor states in the conventional basis. (a)~$S^{(2)}$ vs subsystem size $l$ for all states with $N=9$. The color denotes $\langle S^2_{\text{tot}}\rangle$ of the corresponding state. Example states showcasing the (b)~area-law, (c)~log-law, and (d)~volume-law entanglement scaling of half-chain entropy. Lines in (c) are fits to $a\ln N+b$.}
\label{fig:conventional}
\end{figure}

We focus mainly on the R\'enyi-2 entanglement entropy $S^{(2)}$ for the symmetric tensor state, because unlike with the von Neumann entropy, $S^{\text{vN}}$, we have developed statistical mechanical techniques for its efficient computation and analytic treatment~\cite{SuppMat} -- enabling us to deduce the entanglement scaling.
We are particularly interested in the scaling behavior with system size, and whether it is thermal or athermal.
$S^{\text{vN}}$ is found to follow similar behavior as $S^{(2)}$ for smaller system sizes~\cite{SuppMat}.
We find that the half-chain $S^{(2)}$ assumes a range of scaling behaviors, summarized in \Cref{fig:Bell} for the Bell basis and \Cref{fig:conventional} for the conventional basis. In addition to the volume law, we find examples of log-law (both for the Bell and conventional bases) and area-law (for conventional basis only) states.

In the Bell basis, we find 
a precise rule to determine which states have volume law or log law entanglement.
We represent a scaling family of states by a triplet ($k_X, k_Y, k_Z$) by setting $n_\alpha = k_\alpha N$.
Whenever the largest $k_\alpha$ is greater than the sum of the other two, the states exhibit volume law, and otherwise, log law [see the phase diagram in \Cref{fig:Bell}(b)].
In \Cref{fig:Bell}(c-e), we show how the entropy changes along the red line marked on the diagram.
We examine $S^{(2)}$ rescaled by (c) $\ln N$ and (d) $N$ to show that the scaling is indeed logarithmic and volume in the two phases.
In \Cref{fig:Bell}(e), we show the susceptibility, $\chi = (1/N)\mathrm{d}^2 S^{(2)} / \mathrm{d} k_z^2$ (identifying $S^{(2)}$ as related to the free energy of a statistical mechanical model~\cite{bao_2020,zhou_nahum_2018,jian_2020}), which exhibits a crossing indicating a second-order entanglement phase transition (akin to famous examples in localized models~\cite{MBL1,ET1} and monitored systems~\cite{ET2,ET3,ET4,ET5}).

In the conventional basis, the entanglement ranges from zero to maximal [see \Cref{fig:conventional}(a)], and we find families of states covering the entire spectrum of entanglement scaling behaviors, including area, log, and volume law.
An example of area law is shown in \Cref{fig:conventional}(b), where the states ($N{-}n_Z, 0, n_Z$) with fixed $\nz$ have entanglement bounded by a constant, $\ln[(\sqrt{\pi} n_Z!) / \Gamma(n_Z+1/2)]$.
States of type ($N{-}n_-,n_-,0)$ with fixed $n_-$ exhibit logarithmic entanglement [see \Cref{fig:conventional}(c)], following the form $S^{(2)} = n_-\ln (N) - \ln n_-!$ (the $n_-=1$ case was previously studied by us in~\cite{Turner2024Jul}).
Some volume-law states have a counterintuitive behavior, for example, the state $(1,1,1)N/3$ shows a volume-law scaling with a small coefficient but the entanglement grows faster when we increase the proportion of the two product triplets [see \Cref{fig:conventional}(d)].
Such anomalous growth of entanglement originates from the symmetric superposition.
We find the entanglement scaling to become volume-law when both the product triplet numbers $\{n_+, n_-\}$ scale extensively, e.g.\@ the state $(N{\pm}1,N{\mp}1,0)/2$ has $S^{(2)}\approx N\ln 2 - 1/(N+1)$, and hence is maximally entangled in the thermodynamic limit. This demonstrates how the symmetrization of triplet states can generate a hierarchy of entanglement scalings which exemplifies the rich structure of the space of symmetric superpositions.

\section{Local observables and correlation functions}
We now look at the behavior of local observables and whether they imply a thermal behavior of symmetric tensor states.
Expectation values of single-site observables are zero in the Bell basis, $\langle S^{\alpha}_i\rangle=0$, since every site is maximally entangled with the rest of the system in all states. This matches the infinite-temperature average of these quantities and one may be tempted to conclude these states as thermal. However, there are subtleties depending on the range of observables due to the following reasons. First, the root states \Cref{eq:singlet} are product states for specific non-local and noncontiguous bipartitions, and consequently are strongly atypical.
Secondly, antipodal correlators assume nonthermal values, such as $\mathcal{C}[N]=1/4$ (where $\mathcal{C}[l]=\langle {\bf S}_i\cdot{\bf S}_{i+l}\rangle$) for all the symmetric tensor states. Conventionally, the thermal nature of a state is probed via local correlation functions while $\mathcal{C}[N]$ is non-local, and thus experimentally inaccessible in thermodynamically large systems, though within the setting of local operation and classical communication (LOCC)~\cite{Chitambar2014} their athermal character can be probed. Furthermore, we find that at any finite system size, $\mathcal{C}[l<N]$ is a nonzero constant independent of $l$ (except for the root states). Hence, all such states are long-range ordered~\cite{Huse2013Jul, MagnonScar, Desaules2022Jul} and atypical at any finite system size. As a consequence, $\mathcal{C}[l]$ is a single-variable function of $\langle S^2_{\text{tot}}\rangle$, increasing linearly with it, $\mathcal{C}[l]=\langle S^2_{\text{tot}}\rangle/[4N(N-1)]-1/(2N-2)$.

The scaling of local correlation functions with system size is crucial for the ultimate fate of their thermal nature. In the Bell basis, we obtain $\mathcal{C}[l] = (n_X n_Y + n_Y n_Z + n_Z n_X)/(2N^2-2N)$. This vanishes as $1/N$ for states with one of the $n_i$ scaling as $N$ while the other two are constant. However, as soon as two of the three $n_i$ scale with $N$, $\mathcal{C}[l]$ assumes a nonzero constant value even in the thermodynamic limit. For example, $\mathcal{C}[l]$ for the states ($2N/3, N/3, 0$) and ($N/3, N/3, N/3$) approaches $1/9$ and $1/6$, respectively~\cite{SuppMat}, in the thermodynamic limit; this proves (even in the conventional sense) that these states are nonthermal. Interestingly, though the latter state follows logarithmic entanglement scaling in agreement with the behavior of many atypical states, the former exhibits volume law and hence can be referred to as exceptional QMBS with volume-law entanglement, going beyond the previously studied cases of volume-law scars~\cite{EAP,Mohapatra2024Oct}.

In the conventional basis, the correlations are given by $\mathcal{C}[l]=((n_+-n_-)^2+(n_++n_-)(2n_Z-1))/(2N(2N-2))$. Due to the presence of product states in this basis, $\mathcal{C}[l]$ is generically nonthermal $(\mathcal{C}[l]\neq 0)$ even in the thermodynamic limit.
There are two limiting cases when correlations exhibit thermal behavior: (1) $n_+$, $n_-$ are constant, and (2) $n_+$, $n_-$ are both extensive and differ by $O(1)$ with a constant $n_Z$.
While both cases have volume-law entanglement, the former belongs to the same class as the entangled-antipodal-pair states, while the latter has a more complex structure. Note that $\mathcal{C}[l]$ can be negative for some states (with $n_Z=0$ and $|n_+-n_-|<\sqrt{N}$) in this basis for any finite system size. To summarize, we have shown that the symmetric tensor states in general exhibit nonthermal behavior in their correlations.

\section{Quasiparticle excitations and asymptotic QMBS}
We consider quasiparticle excitations \cite{Moudgalya_2018_2, Lin_2019, MagnonScar, quasiparticle_scar_review} on top of the singlet root state, generated by a local operator $Q_j=\sigma_{\alpha,j}$, where $\alpha \in \{X,Y,Z,{+},{-}\}$, the action of which on one singlet is to replace it with the appropriate triplet.
We find $S^2_\text{tot} Q_j\ket{S} = 2 Q_j\ket{S}$, hence this excitation carries spin-1 and should be referred to as a triplon.
We can take linear combinations of these operators to find a single-particle wavefunction that represents an asymptotically stable excitation (or asymptotic QMBS~\cite{asymptotic}),
\begin{equation}
    \ket{\mathcal{Q}}=\frac{1}{\sqrt{N}}\sum_{j=1}^N Q_j\ket{S}
    \text{.}
    \label{eq:quasi}
\end{equation}
This wavefunction is a square wave as the superposition is on only half the system.
We find, $E_{\mathcal{Q}}=\bra{\mathcal{Q}} H \ket{\mathcal{Q}}=0$ and the energy variance $\delta = \bra{\mathcal{Q}} H^2 \ket{\mathcal{Q}} - E_{\mathcal{Q}}^2 = 2/N$ (for $r=1$).
Therefore, although $\ket{\mathcal{Q}}$ is not an eigenstate of $H$ at any finite system size, the lifetime of the quasiparticle diverges in the thermodynamic limit.
This can be understood intuitively as follows: the energy fluctuations are proportional to the gradient of the wavefunction, so the most stable excitations will vary spatially as little as possible.
The antisymmetric nature of the singlet state, however, demands that the wavefunction be odd under translation by $N$ as the even-transforming component annihilates $\ket{S}$.
This mandates some minimum amount of variation into the wavefunction.
In the thermodynamic limit, however, the energy fluctuations become negligible because almost all of the wavefunction is far away from the edges of the square wave.
Intriguingly, a family of related quasiparticle eigenstates for this model can be constructed using a generalized Bethe ansatz~\cite{Melendrez2025Jan}.

\section{Extension to higher dimensions}
The state construction relies on the cancellation between the action of a local term in the Hamiltonian with its antipodal partner, while the structure of the rest of the lattice is mostly immaterial. Hence, higher-dimensional generalizations are fairly straightforward, but they also provide a new way to create antipodal pairs. E.g., a rectangular lattice of $2N_x \times 2N_y$ sites (with a specific choice of the staggered interactions and odd $N_x$ and even $N_y$) possesses two antipodal pairs: ($x,y$)--($x + N_x, y + N_y$) and ($x,y$)--($x + N_x, y$). Given this setup, all symmetric tensor states are zero-energy eigenstates of the Hamiltonian. Thus, higher dimensional lattices allow the superposition of multiple ``flavors'' of symmetric states, potentially leading to a liquid-like behavior. Extensions to cubic and hypercubic lattices are similarly straightforward.

\section{Stability against perturbations}
We find that many symmetric tensor scars remain zero energy eigenstates even in the presence of SU(2) symmetry-breaking perturbations such as uniform and staggered fields, or easy-axis anisotropy of the exchange interaction. States that are no longer eigenstates in the presence of such perturbations, remain stable up to significantly long times for moderate values of the perturbations while others exhibit coherent oscillations in the presence of an additional staggered field (see~\cite{SuppMat}). 

\section{Discussion}
In this Letter, we have shown the existence of a class of exact quantum many-body scars at infinite temperature in a family of models with staggered Heisenberg interaction. These symmetric tensor scars are constructed by long-range triplet coverings of the system. Some of these (root) states exhibit a mixture of thermal (i.e.\@ volume-law scaling of entanglement and thermal values of local observables) and nonthermal (product structure for certain bipartite entanglement cuts, strictly nonthermal antipodal correlation) properties. We demonstrate how to regularize such anomalous QMBS by inducing a non-thermal local expectation value via symmetric superposition of different triplet coverings. We note that the relationship between the frame of root states and the symmetric tensor scars mirrors the relationship between the mean-field TDVP frame~\cite{Ho:2019} used to study scar states in the PXP model and the permutation-invariant scar quasimodes~\cite{Turner:2021} which result from the quantization of that semiclassical frame. The ergodic properties of the states are sensitive to the choice of the triplet basis. While the Bell pair basis admits log and volume-law scaling but no area-law states, the conventional basis supports the full spectrum of entanglement behaviors. 
Many of these states are also stable against symmetry-breaking perturbations~\cite{SuppMat}.
Quasiparticles on top of the singlet root state are found to be asymptotic QMBS. We also discuss the recipe for generalization to higher dimensions.

Our results open up numerous diverse research directions.
First and foremost, our recipe for creating highly entangled states unlocks the way of building exact many-body scars beyond the usual area-law paradigm. In fact, since in the conventional triplet basis, the states admit any behavior from area to volume law, one can construct states with exotic entanglement scaling [such as $\sim \ln^2 N$ or a fractal entanglement $\sim N^\alpha, \alpha \in (0,1)$]. This allows analytical insights into unusual families of states, such as those exhibiting quantum critical properties or multifractality. 
Additionally, similar symmetric tensor constructions are viable, such as those involving a superposition of three or more spin state coverings, although whether such constructions would lead to new eigenstates is an open question.
Beyond this, the possibility of easily creating scars in higher-dimensional systems raises the question of whether one can produce topological symmetric tensor scars, potentially providing analytical leverage on non-trivial anyonic excitations (extending our work on quasiparticles) and other topological phenomena.
These novel state construction techniques can not only lead to theoretical insights into complex dynamical properties, but also provide a framework for stabilizing quantum order in thermal systems.
The simulation of such symmetric tensor scars in near-term quantum computers is also an interesting future avenue to explore~\cite{Chen2022Oct,Gustafson2023Nov,Dong2023Dec}.

\begin{acknowledgments}
\textit{Acknowledgments.}---%
We thank Hitesh J. Changlani and Ronald Melendrez for discussions as well as collaboration on related projects.
A.~P.,  B.~M., and M.~S.\@ were funded by the European Research Council (ERC) under the European Union's Horizon 2020 research and innovation programme (Grant Agreement No.\@ 853368). C.~J.~T.\@ is supported by an EPSRC fellowship (Grant Ref. EP/W005743/1).
B.~M.\@ was also funded by DST, Government of India via the INSPIRE Faculty programme.
M.~S. was also supported by the Engineering and Physical Sciences Research Council (EPSRC) grant on Robust and Reliable Quantum Computing (RoaRQ), Investigation 004 (grant reference EP/W032635/1), and the EPSRC grant on (De)constructing quantum software (DeQS) (grant reference EP/Z002230/1).
The authors acknowledge the use of the UCL High Performance Computing Facilities (Myriad and Kathleen), and associated support services, in the completion of this work.
\end{acknowledgments}

\textit{Data availability.}---The data that support the findings of this article are openly available~\cite{Data}.

\bibliography{refs}

\begin{thebibliography}{81}%
\makeatletter
\providecommand \@ifxundefined [1]{%
 \@ifx{#1\undefined}
}%
\providecommand \@ifnum [1]{%
 \ifnum #1\expandafter \@firstoftwo
 \else \expandafter \@secondoftwo
 \fi
}%
\providecommand \@ifx [1]{%
 \ifx #1\expandafter \@firstoftwo
 \else \expandafter \@secondoftwo
 \fi
}%
\providecommand \natexlab [1]{#1}%
\providecommand \enquote  [1]{``#1''}%
\providecommand \bibnamefont  [1]{#1}%
\providecommand \bibfnamefont [1]{#1}%
\providecommand \citenamefont [1]{#1}%
\providecommand \href@noop [0]{\@secondoftwo}%
\providecommand \href [0]{\begingroup \@sanitize@url \@href}%
\providecommand \@href[1]{\@@startlink{#1}\@@href}%
\providecommand \@@href[1]{\endgroup#1\@@endlink}%
\providecommand \@sanitize@url [0]{\catcode `\\12\catcode `\$12\catcode `\&12\catcode `\#12\catcode `\^12\catcode `\_12\catcode `\%12\relax}%
\providecommand \@@startlink[1]{}%
\providecommand \@@endlink[0]{}%
\providecommand \url  [0]{\begingroup\@sanitize@url \@url }%
\providecommand \@url [1]{\endgroup\@href {#1}{\urlprefix }}%
\providecommand \urlprefix  [0]{URL }%
\providecommand \Eprint [0]{\href }%
\providecommand \doibase [0]{https://doi.org/}%
\providecommand \selectlanguage [0]{\@gobble}%
\providecommand \bibinfo  [0]{\@secondoftwo}%
\providecommand \bibfield  [0]{\@secondoftwo}%
\providecommand \translation [1]{[#1]}%
\providecommand \BibitemOpen [0]{}%
\providecommand \bibitemStop [0]{}%
\providecommand \bibitemNoStop [0]{.\EOS\space}%
\providecommand \EOS [0]{\spacefactor3000\relax}%
\providecommand \BibitemShut  [1]{\csname bibitem#1\endcsname}%
\let\auto@bib@innerbib\@empty
\bibitem [{\citenamefont {Rigol}\ \emph {et~al.}(2008)\citenamefont {Rigol}, \citenamefont {Dunjko},\ and\ \citenamefont {Olshanii}}]{Rigol2008}%
  \BibitemOpen
  \bibfield  {author} {\bibinfo {author} {\bibfnamefont {M.}~\bibnamefont {Rigol}}, \bibinfo {author} {\bibfnamefont {V.}~\bibnamefont {Dunjko}},\ and\ \bibinfo {author} {\bibfnamefont {M.}~\bibnamefont {Olshanii}},\ }\href {https://doi.org/10.1038/nature06838} {\bibfield  {journal} {\bibinfo  {journal} {Nature}\ }\textbf {\bibinfo {volume} {452}},\ \bibinfo {pages} {854} (\bibinfo {year} {2008})}\BibitemShut {NoStop}%
\bibitem [{\citenamefont {D'Alessio}\ \emph {et~al.}(2016)\citenamefont {D'Alessio}, \citenamefont {Kafri}, \citenamefont {Polkovnikov},\ and\ \citenamefont {Rigol}}]{DAlessio2016}%
  \BibitemOpen
  \bibfield  {author} {\bibinfo {author} {\bibfnamefont {L.}~\bibnamefont {D'Alessio}}, \bibinfo {author} {\bibfnamefont {Y.}~\bibnamefont {Kafri}}, \bibinfo {author} {\bibfnamefont {A.}~\bibnamefont {Polkovnikov}},\ and\ \bibinfo {author} {\bibfnamefont {M.}~\bibnamefont {Rigol}},\ }\href {https://doi.org/10.1080/00018732.2016.1198134} {\bibfield  {journal} {\bibinfo  {journal} {Adv. Phys.}\ }\textbf {\bibinfo {volume} {65}},\ \bibinfo {pages} {239} (\bibinfo {year} {2016})}\BibitemShut {NoStop}%
\bibitem [{\citenamefont {Kim}\ \emph {et~al.}(2014)\citenamefont {Kim}, \citenamefont {Ikeda},\ and\ \citenamefont {Huse}}]{strongETH}%
  \BibitemOpen
  \bibfield  {author} {\bibinfo {author} {\bibfnamefont {H.}~\bibnamefont {Kim}}, \bibinfo {author} {\bibfnamefont {T.~N.}\ \bibnamefont {Ikeda}},\ and\ \bibinfo {author} {\bibfnamefont {D.~A.}\ \bibnamefont {Huse}},\ }\href {https://doi.org/10.1103/PhysRevE.90.052105} {\bibfield  {journal} {\bibinfo  {journal} {Phys. Rev. E}\ }\textbf {\bibinfo {volume} {90}},\ \bibinfo {pages} {052105} (\bibinfo {year} {2014})}\BibitemShut {NoStop}%
\bibitem [{\citenamefont {Turner}\ \emph {et~al.}(2018{\natexlab{a}})\citenamefont {Turner}, \citenamefont {Michailidis}, \citenamefont {Abanin}, \citenamefont {Serbyn},\ and\ \citenamefont {Papi{\ifmmode\acute{c}\else\'{c}\fi}}}]{Turner:2018a}%
  \BibitemOpen
  \bibfield  {author} {\bibinfo {author} {\bibfnamefont {C.~J.}\ \bibnamefont {Turner}}, \bibinfo {author} {\bibfnamefont {A.~A.}\ \bibnamefont {Michailidis}}, \bibinfo {author} {\bibfnamefont {D.~A.}\ \bibnamefont {Abanin}}, \bibinfo {author} {\bibfnamefont {M.}~\bibnamefont {Serbyn}},\ and\ \bibinfo {author} {\bibfnamefont {Z.}~\bibnamefont {Papi{\ifmmode\acute{c}\else\'{c}\fi}}},\ }\href {https://doi.org/10.1038/s41567-018-0137-5} {\bibfield  {journal} {\bibinfo  {journal} {Nat. Phys.}\ }\textbf {\bibinfo {volume} {14}},\ \bibinfo {pages} {745} (\bibinfo {year} {2018}{\natexlab{a}})}\BibitemShut {NoStop}%
\bibitem [{\citenamefont {Turner}\ \emph {et~al.}(2018{\natexlab{b}})\citenamefont {Turner}, \citenamefont {Michailidis}, \citenamefont {Abanin}, \citenamefont {Serbyn},\ and\ \citenamefont {Papi{\ifmmode\acute{c}\else\'{c}\fi}}}]{Turner:2018b}%
  \BibitemOpen
  \bibfield  {author} {\bibinfo {author} {\bibfnamefont {C.~J.}\ \bibnamefont {Turner}}, \bibinfo {author} {\bibfnamefont {A.~A.}\ \bibnamefont {Michailidis}}, \bibinfo {author} {\bibfnamefont {D.~A.}\ \bibnamefont {Abanin}}, \bibinfo {author} {\bibfnamefont {M.}~\bibnamefont {Serbyn}},\ and\ \bibinfo {author} {\bibfnamefont {Z.}~\bibnamefont {Papi{\ifmmode\acute{c}\else\'{c}\fi}}},\ }\href {https://doi.org/10.1103/PhysRevB.98.155134} {\bibfield  {journal} {\bibinfo  {journal} {Phys. Rev. B}\ }\textbf {\bibinfo {volume} {98}},\ \bibinfo {pages} {155134} (\bibinfo {year} {2018}{\natexlab{b}})}\BibitemShut {NoStop}%
\bibitem [{\citenamefont {Moudgalya}\ \emph {et~al.}(2018)\citenamefont {Moudgalya}, \citenamefont {Rachel}, \citenamefont {Bernevig},\ and\ \citenamefont {Regnault}}]{Moudgalya_2018_2}%
  \BibitemOpen
  \bibfield  {author} {\bibinfo {author} {\bibfnamefont {S.}~\bibnamefont {Moudgalya}}, \bibinfo {author} {\bibfnamefont {S.}~\bibnamefont {Rachel}}, \bibinfo {author} {\bibfnamefont {B.~A.}\ \bibnamefont {Bernevig}},\ and\ \bibinfo {author} {\bibfnamefont {N.}~\bibnamefont {Regnault}},\ }\href {https://doi.org/10.1103/PhysRevB.98.235155} {\bibfield  {journal} {\bibinfo  {journal} {Phys. Rev. B}\ }\textbf {\bibinfo {volume} {98}},\ \bibinfo {pages} {235155} (\bibinfo {year} {2018})}\BibitemShut {NoStop}%
\bibitem [{\citenamefont {Lin}\ and\ \citenamefont {Motrunich}(2019)}]{Lin_2019}%
  \BibitemOpen
  \bibfield  {author} {\bibinfo {author} {\bibfnamefont {C.-J.}\ \bibnamefont {Lin}}\ and\ \bibinfo {author} {\bibfnamefont {O.~I.}\ \bibnamefont {Motrunich}},\ }\href {https://doi.org/10.1103/PhysRevLett.122.173401} {\bibfield  {journal} {\bibinfo  {journal} {Phys. Rev. Lett.}\ }\textbf {\bibinfo {volume} {122}},\ \bibinfo {pages} {173401} (\bibinfo {year} {2019})}\BibitemShut {NoStop}%
\bibitem [{\citenamefont {Iadecola}\ and\ \citenamefont {Schecter}(2020)}]{Iadecola2020}%
  \BibitemOpen
  \bibfield  {author} {\bibinfo {author} {\bibfnamefont {T.}~\bibnamefont {Iadecola}}\ and\ \bibinfo {author} {\bibfnamefont {M.}~\bibnamefont {Schecter}},\ }\href {https://doi.org/10.1103/PhysRevB.101.024306} {\bibfield  {journal} {\bibinfo  {journal} {Phys. Rev. B}\ }\textbf {\bibinfo {volume} {101}},\ \bibinfo {pages} {024306} (\bibinfo {year} {2020})}\BibitemShut {NoStop}%
\bibitem [{\citenamefont {Lee}\ \emph {et~al.}(2020)\citenamefont {Lee}, \citenamefont {Melendrez}, \citenamefont {Pal},\ and\ \citenamefont {Changlani}}]{Lee2020}%
  \BibitemOpen
  \bibfield  {author} {\bibinfo {author} {\bibfnamefont {K.}~\bibnamefont {Lee}}, \bibinfo {author} {\bibfnamefont {R.}~\bibnamefont {Melendrez}}, \bibinfo {author} {\bibfnamefont {A.}~\bibnamefont {Pal}},\ and\ \bibinfo {author} {\bibfnamefont {H.~J.}\ \bibnamefont {Changlani}},\ }\href {https://doi.org/10.1103/PhysRevB.101.241111} {\bibfield  {journal} {\bibinfo  {journal} {Phys. Rev. B}\ }\textbf {\bibinfo {volume} {101}},\ \bibinfo {pages} {241111} (\bibinfo {year} {2020})}\BibitemShut {NoStop}%
\bibitem [{\citenamefont {Wildeboer}\ \emph {et~al.}(2021)\citenamefont {Wildeboer}, \citenamefont {Seidel}, \citenamefont {Srivatsa}, \citenamefont {Nielsen},\ and\ \citenamefont {Erten}}]{Wildeboer2021}%
  \BibitemOpen
  \bibfield  {author} {\bibinfo {author} {\bibfnamefont {J.}~\bibnamefont {Wildeboer}}, \bibinfo {author} {\bibfnamefont {A.}~\bibnamefont {Seidel}}, \bibinfo {author} {\bibfnamefont {N.~S.}\ \bibnamefont {Srivatsa}}, \bibinfo {author} {\bibfnamefont {A.~E.~B.}\ \bibnamefont {Nielsen}},\ and\ \bibinfo {author} {\bibfnamefont {O.}~\bibnamefont {Erten}},\ }\href {https://doi.org/10.1103/PhysRevB.104.L121103} {\bibfield  {journal} {\bibinfo  {journal} {Phys. Rev. B}\ }\textbf {\bibinfo {volume} {104}},\ \bibinfo {pages} {L121103} (\bibinfo {year} {2021})}\BibitemShut {NoStop}%
\bibitem [{\citenamefont {Srivatsa}\ \emph {et~al.}(2020)\citenamefont {Srivatsa}, \citenamefont {Wildeboer}, \citenamefont {Seidel},\ and\ \citenamefont {Nielsen}}]{Srivatsa2020}%
  \BibitemOpen
  \bibfield  {author} {\bibinfo {author} {\bibfnamefont {N.~S.}\ \bibnamefont {Srivatsa}}, \bibinfo {author} {\bibfnamefont {J.}~\bibnamefont {Wildeboer}}, \bibinfo {author} {\bibfnamefont {A.}~\bibnamefont {Seidel}},\ and\ \bibinfo {author} {\bibfnamefont {A.~E.~B.}\ \bibnamefont {Nielsen}},\ }\href {https://doi.org/10.1103/PhysRevB.102.235106} {\bibfield  {journal} {\bibinfo  {journal} {Phys. Rev. B}\ }\textbf {\bibinfo {volume} {102}},\ \bibinfo {pages} {235106} (\bibinfo {year} {2020})}\BibitemShut {NoStop}%
\bibitem [{\citenamefont {Serbyn}\ \emph {et~al.}(2021)\citenamefont {Serbyn}, \citenamefont {Abanin},\ and\ \citenamefont {Papi{\ifmmode\acute{c}\else\'{c}\fi}}}]{serbyn2021QMBS}%
  \BibitemOpen
  \bibfield  {author} {\bibinfo {author} {\bibfnamefont {M.}~\bibnamefont {Serbyn}}, \bibinfo {author} {\bibfnamefont {D.~A.}\ \bibnamefont {Abanin}},\ and\ \bibinfo {author} {\bibfnamefont {Z.}~\bibnamefont {Papi{\ifmmode\acute{c}\else\'{c}\fi}}},\ }\href {https://doi.org/10.1038/s41567-021-01230-2} {\bibfield  {journal} {\bibinfo  {journal} {Nat. Phys.}\ }\textbf {\bibinfo {volume} {17}},\ \bibinfo {pages} {675} (\bibinfo {year} {2021})}\BibitemShut {NoStop}%
\bibitem [{\citenamefont {Richter}\ and\ \citenamefont {Pal}(2022)}]{Richter2022}%
  \BibitemOpen
  \bibfield  {author} {\bibinfo {author} {\bibfnamefont {J.}~\bibnamefont {Richter}}\ and\ \bibinfo {author} {\bibfnamefont {A.}~\bibnamefont {Pal}},\ }\href {https://doi.org/10.1103/PhysRevResearch.4.L012003} {\bibfield  {journal} {\bibinfo  {journal} {Phys. Rev. Res.}\ }\textbf {\bibinfo {volume} {4}},\ \bibinfo {pages} {L012003} (\bibinfo {year} {2022})}\BibitemShut {NoStop}%
\bibitem [{\citenamefont {Sanjay~Moudgalya}\ and\ \citenamefont {Regnault}(2022)}]{sanjay_review}%
  \BibitemOpen
  \bibfield  {author} {\bibinfo {author} {\bibfnamefont {B.~A.~B.}\ \bibnamefont {Sanjay~Moudgalya}}\ and\ \bibinfo {author} {\bibfnamefont {N.}~\bibnamefont {Regnault}},\ }\href {https://doi.org/10.1088/1361-6633/ac73a0} {\bibfield  {journal} {\bibinfo  {journal} {Rep. Prog. Phys.}\ }\textbf {\bibinfo {volume} {85}},\ \bibinfo {pages} {086501} (\bibinfo {year} {2022})}\BibitemShut {NoStop}%
\bibitem [{\citenamefont {Chandran}\ \emph {et~al.}(2023)\citenamefont {Chandran}, \citenamefont {Iadecola}, \citenamefont {Khemani},\ and\ \citenamefont {Moessner}}]{quasiparticle_scar_review}%
  \BibitemOpen
  \bibfield  {author} {\bibinfo {author} {\bibfnamefont {A.}~\bibnamefont {Chandran}}, \bibinfo {author} {\bibfnamefont {T.}~\bibnamefont {Iadecola}}, \bibinfo {author} {\bibfnamefont {V.}~\bibnamefont {Khemani}},\ and\ \bibinfo {author} {\bibfnamefont {R.}~\bibnamefont {Moessner}},\ }\href {https://doi.org/10.1146/annurev-conmatphys-031620-101617} {\bibfield  {journal} {\bibinfo  {journal} {Annu. Rev. Condens. Matter Phys.}\ }\textbf {\bibinfo {volume} {14}},\ \bibinfo {pages} {443} (\bibinfo {year} {2023})}\BibitemShut {NoStop}%
\bibitem [{\citenamefont {Bernien}\ \emph {et~al.}(2017)\citenamefont {Bernien}, \citenamefont {Schwartz}, \citenamefont {Keesling}, \citenamefont {Levine}, \citenamefont {Omran}, \citenamefont {Pichler}, \citenamefont {Choi}, \citenamefont {Zibrov}, \citenamefont {Endres}, \citenamefont {Greiner}, \citenamefont {Vuletic},\ and\ \citenamefont {Lukin}}]{Bernien2017}%
  \BibitemOpen
  \bibfield  {author} {\bibinfo {author} {\bibfnamefont {H.}~\bibnamefont {Bernien}}, \bibinfo {author} {\bibfnamefont {S.}~\bibnamefont {Schwartz}}, \bibinfo {author} {\bibfnamefont {A.}~\bibnamefont {Keesling}}, \bibinfo {author} {\bibfnamefont {H.}~\bibnamefont {Levine}}, \bibinfo {author} {\bibfnamefont {A.}~\bibnamefont {Omran}}, \bibinfo {author} {\bibfnamefont {H.}~\bibnamefont {Pichler}}, \bibinfo {author} {\bibfnamefont {S.}~\bibnamefont {Choi}}, \bibinfo {author} {\bibfnamefont {A.~S.}\ \bibnamefont {Zibrov}}, \bibinfo {author} {\bibfnamefont {M.}~\bibnamefont {Endres}}, \bibinfo {author} {\bibfnamefont {M.}~\bibnamefont {Greiner}}, \bibinfo {author} {\bibfnamefont {V.}~\bibnamefont {Vuletic}},\ and\ \bibinfo {author} {\bibfnamefont {M.~D.}\ \bibnamefont {Lukin}},\ }\href {https://doi.org/10.1038/nature24622} {\bibfield  {journal} {\bibinfo  {journal} {Nature}\ }\textbf {\bibinfo {volume} {551}},\ \bibinfo {pages} {579} (\bibinfo {year} {2017})}\BibitemShut {NoStop}%
\bibitem [{\citenamefont {Jeyaretnam}\ \emph {et~al.}(2021)\citenamefont {Jeyaretnam}, \citenamefont {Richter},\ and\ \citenamefont {Pal}}]{Jeyaretnam2021Jul}%
  \BibitemOpen
  \bibfield  {author} {\bibinfo {author} {\bibfnamefont {J.}~\bibnamefont {Jeyaretnam}}, \bibinfo {author} {\bibfnamefont {J.}~\bibnamefont {Richter}},\ and\ \bibinfo {author} {\bibfnamefont {A.}~\bibnamefont {Pal}},\ }\href {https://doi.org/10.1103/PhysRevB.104.014424} {\bibfield  {journal} {\bibinfo  {journal} {Phys. Rev. B}\ }\textbf {\bibinfo {volume} {104}},\ \bibinfo {pages} {014424} (\bibinfo {year} {2021})}\BibitemShut {NoStop}%
\bibitem [{\citenamefont {Shiraishi}\ and\ \citenamefont {Mori}(2017)}]{Shiraishi:2017}%
  \BibitemOpen
  \bibfield  {author} {\bibinfo {author} {\bibfnamefont {N.}~\bibnamefont {Shiraishi}}\ and\ \bibinfo {author} {\bibfnamefont {T.}~\bibnamefont {Mori}},\ }\href {https://doi.org/10.1103/physrevlett.119.030601} {\bibfield  {journal} {\bibinfo  {journal} {Phys. Rev. Lett.}\ }\textbf {\bibinfo {volume} {119}},\ \bibinfo {pages} {030601} (\bibinfo {year} {2017})}\BibitemShut {NoStop}%
\bibitem [{\citenamefont {Langlett}\ \emph {et~al.}(2022)\citenamefont {Langlett}, \citenamefont {Yang}, \citenamefont {Wildeboer}, \citenamefont {Gorshkov}, \citenamefont {Iadecola},\ and\ \citenamefont {Xu}}]{rainbow_scars}%
  \BibitemOpen
  \bibfield  {author} {\bibinfo {author} {\bibfnamefont {C.~M.}\ \bibnamefont {Langlett}}, \bibinfo {author} {\bibfnamefont {Z.-C.}\ \bibnamefont {Yang}}, \bibinfo {author} {\bibfnamefont {J.}~\bibnamefont {Wildeboer}}, \bibinfo {author} {\bibfnamefont {A.~V.}\ \bibnamefont {Gorshkov}}, \bibinfo {author} {\bibfnamefont {T.}~\bibnamefont {Iadecola}},\ and\ \bibinfo {author} {\bibfnamefont {S.}~\bibnamefont {Xu}},\ }\href {https://doi.org/10.1103/PhysRevB.105.L060301} {\bibfield  {journal} {\bibinfo  {journal} {Phys. Rev. B}\ }\textbf {\bibinfo {volume} {105}},\ \bibinfo {pages} {L060301} (\bibinfo {year} {2022})}\BibitemShut {NoStop}%
\bibitem [{\citenamefont {Vitagliano}\ \emph {et~al.}(2010)\citenamefont {Vitagliano}, \citenamefont {Riera},\ and\ \citenamefont {Latorre}}]{rainbow1}%
  \BibitemOpen
  \bibfield  {author} {\bibinfo {author} {\bibfnamefont {G.}~\bibnamefont {Vitagliano}}, \bibinfo {author} {\bibfnamefont {A.}~\bibnamefont {Riera}},\ and\ \bibinfo {author} {\bibfnamefont {J.~I.}\ \bibnamefont {Latorre}},\ }\href {https://doi.org/10.1088/1367-2630/12/11/113049} {\bibfield  {journal} {\bibinfo  {journal} {New J. Phys.}\ }\textbf {\bibinfo {volume} {12}},\ \bibinfo {pages} {113049} (\bibinfo {year} {2010})}\BibitemShut {NoStop}%
\bibitem [{\citenamefont {Ram{\ifmmode\acute{\imath}\else\'{\i}\fi}rez}\ \emph {et~al.}(2014)\citenamefont {Ram{\ifmmode\acute{\imath}\else\'{\i}\fi}rez}, \citenamefont {Rodr{\ifmmode\acute{\imath}\else\'{\i}\fi}guez-Laguna},\ and\ \citenamefont {Sierra}}]{rainbow2}%
  \BibitemOpen
  \bibfield  {author} {\bibinfo {author} {\bibfnamefont {G.}~\bibnamefont {Ram{\ifmmode\acute{\imath}\else\'{\i}\fi}rez}}, \bibinfo {author} {\bibfnamefont {J.}~\bibnamefont {Rodr{\ifmmode\acute{\imath}\else\'{\i}\fi}guez-Laguna}},\ and\ \bibinfo {author} {\bibfnamefont {G.}~\bibnamefont {Sierra}},\ }\href {https://doi.org/10.1088/1742-5468/2014/10/P10004} {\bibfield  {journal} {\bibinfo  {journal} {J. Stat. Mech.: Theory Exp.}\ }\textbf {\bibinfo {volume} {2014}}\bibinfo  {number} { (10)},\ \bibinfo {pages} {P10004}}\BibitemShut {NoStop}%
\bibitem [{\citenamefont {Ram{\ifmmode\acute{\imath}\else\'{\i}\fi}rez}\ \emph {et~al.}(2015)\citenamefont {Ram{\ifmmode\acute{\imath}\else\'{\i}\fi}rez}, \citenamefont {Rodr{\ifmmode\acute{\imath}\else\'{\i}\fi}guez-Laguna},\ and\ \citenamefont {Sierra}}]{rainbow3}%
  \BibitemOpen
\bibfield  {number} {  }\bibfield  {author} {\bibinfo {author} {\bibfnamefont {G.}~\bibnamefont {Ram{\ifmmode\acute{\imath}\else\'{\i}\fi}rez}}, \bibinfo {author} {\bibfnamefont {J.}~\bibnamefont {Rodr{\ifmmode\acute{\imath}\else\'{\i}\fi}guez-Laguna}},\ and\ \bibinfo {author} {\bibfnamefont {G.}~\bibnamefont {Sierra}},\ }\href {https://doi.org/10.1088/1742-5468/2015/06/P06002} {\bibfield  {journal} {\bibinfo  {journal} {J. Stat. Mech.: Theory Exp.}\ }\textbf {\bibinfo {volume} {2015}}\bibinfo  {number} { (6)},\ \bibinfo {pages} {P06002}}\BibitemShut {NoStop}%
\bibitem [{\citenamefont {Maldacena}(2003)}]{Juan_Maldacena_2003}%
  \BibitemOpen
\bibfield  {number} {  }\bibfield  {author} {\bibinfo {author} {\bibfnamefont {J.}~\bibnamefont {Maldacena}},\ }\href {https://doi.org/10.1088/1126-6708/2003/04/021} {\bibfield  {journal} {\bibinfo  {journal} {J. High Energy Phys.}\ }\textbf {\bibinfo {volume} {2003}}\bibinfo  {number} { (04)},\ \bibinfo {pages} {021}}\BibitemShut {NoStop}%
\bibitem [{\citenamefont {Hartman}\ and\ \citenamefont {Maldacena}(2013)}]{Hartmann}%
  \BibitemOpen
\bibfield  {number} {  }\bibfield  {author} {\bibinfo {author} {\bibfnamefont {T.}~\bibnamefont {Hartman}}\ and\ \bibinfo {author} {\bibfnamefont {J.}~\bibnamefont {Maldacena}},\ }\href {https://doi.org/10.1007/JHEP05(2013)014} {\bibfield  {journal} {\bibinfo  {journal} {J. High Energy Phys.}\ }\textbf {\bibinfo {volume} {2013}}\bibinfo  {number} { (5)},\ \bibinfo {pages} {1}}\BibitemShut {NoStop}%
\bibitem [{\citenamefont {Papadodimas}\ and\ \citenamefont {Raju}(2015)}]{Raju}%
  \BibitemOpen
\bibfield  {number} {  }\bibfield  {author} {\bibinfo {author} {\bibfnamefont {K.}~\bibnamefont {Papadodimas}}\ and\ \bibinfo {author} {\bibfnamefont {S.}~\bibnamefont {Raju}},\ }\href {https://doi.org/10.1103/PhysRevLett.115.211601} {\bibfield  {journal} {\bibinfo  {journal} {Phys. Rev. Lett.}\ }\textbf {\bibinfo {volume} {115}},\ \bibinfo {pages} {211601} (\bibinfo {year} {2015})}\BibitemShut {NoStop}%
\bibitem [{\citenamefont {Cottrell}\ \emph {et~al.}(2019)\citenamefont {Cottrell}, \citenamefont {Freivogel}, \citenamefont {Hofman},\ and\ \citenamefont {Lokhande}}]{Cottrell}%
  \BibitemOpen
  \bibfield  {author} {\bibinfo {author} {\bibfnamefont {W.}~\bibnamefont {Cottrell}}, \bibinfo {author} {\bibfnamefont {B.}~\bibnamefont {Freivogel}}, \bibinfo {author} {\bibfnamefont {D.~M.}\ \bibnamefont {Hofman}},\ and\ \bibinfo {author} {\bibfnamefont {S.~F.}\ \bibnamefont {Lokhande}},\ }\href {https://doi.org/10.1007/JHEP02(2019)058} {\bibfield  {journal} {\bibinfo  {journal} {J. High Energy Phys.}\ }\textbf {\bibinfo {volume} {2019}}\bibinfo  {number} { (2)},\ \bibinfo {pages} {1}}\BibitemShut {NoStop}%
\bibitem [{\citenamefont {Chiba}\ and\ \citenamefont {Yoneta}(2024)}]{EAP}%
  \BibitemOpen
\bibfield  {number} {  }\bibfield  {author} {\bibinfo {author} {\bibfnamefont {Y.}~\bibnamefont {Chiba}}\ and\ \bibinfo {author} {\bibfnamefont {Y.}~\bibnamefont {Yoneta}},\ }\href {https://doi.org/10.1103/PhysRevLett.133.170404} {\bibfield  {journal} {\bibinfo  {journal} {Phys. Rev. Lett.}\ }\textbf {\bibinfo {volume} {133}},\ \bibinfo {pages} {170404} (\bibinfo {year} {2024})}\BibitemShut {NoStop}%
\bibitem [{\citenamefont {Mohapatra}\ \emph {et~al.}(2025)\citenamefont {Mohapatra}, \citenamefont {Moudgalya},\ and\ \citenamefont {Balram}}]{Mohapatra2024Oct}%
  \BibitemOpen
  \bibfield  {author} {\bibinfo {author} {\bibfnamefont {S.}~\bibnamefont {Mohapatra}}, \bibinfo {author} {\bibfnamefont {S.}~\bibnamefont {Moudgalya}},\ and\ \bibinfo {author} {\bibfnamefont {A.~C.}\ \bibnamefont {Balram}},\ }\href {https://doi.org/10.1103/PhysRevLett.134.210403} {\bibfield  {journal} {\bibinfo  {journal} {Phys. Rev. Lett.}\ }\textbf {\bibinfo {volume} {134}},\ \bibinfo {pages} {210403} (\bibinfo {year} {2025})}\BibitemShut {NoStop}%
\bibitem [{\citenamefont {Ivanov}\ and\ \citenamefont {Motrunich}(2025)}]{Ivanov2024Mar}%
  \BibitemOpen
  \bibfield  {author} {\bibinfo {author} {\bibfnamefont {A.~N.}\ \bibnamefont {Ivanov}}\ and\ \bibinfo {author} {\bibfnamefont {O.~I.}\ \bibnamefont {Motrunich}},\ }\href {https://doi.org/10.1103/PhysRevLett.134.050403} {\bibfield  {journal} {\bibinfo  {journal} {Phys. Rev. Lett.}\ }\textbf {\bibinfo {volume} {134}},\ \bibinfo {pages} {050403} (\bibinfo {year} {2025})}\BibitemShut {NoStop}%
\bibitem [{\citenamefont {Caetano}\ and\ \citenamefont {Komatsu}(2022)}]{Crosscap1}%
  \BibitemOpen
  \bibfield  {author} {\bibinfo {author} {\bibfnamefont {J.}~\bibnamefont {Caetano}}\ and\ \bibinfo {author} {\bibfnamefont {S.}~\bibnamefont {Komatsu}},\ }\href {https://doi.org/10.1007/s10955-022-02914-6} {\bibfield  {journal} {\bibinfo  {journal} {J. Stat. Phys.}\ }\textbf {\bibinfo {volume} {187}},\ \bibinfo {pages} {1} (\bibinfo {year} {2022})}\BibitemShut {NoStop}%
\bibitem [{\citenamefont {Gombor}(2022)}]{Gombor2022Jul}%
  \BibitemOpen
  \bibfield  {author} {\bibinfo {author} {\bibfnamefont {T.}~\bibnamefont {Gombor}},\ }\bibfield  {journal} {\bibinfo  {journal} {arXiv}\ }\href {https://doi.org/10.48550/arXiv.2207.10598} {10.48550/arXiv.2207.10598} (\bibinfo {year} {2022}),\ \Eprint {https://arxiv.org/abs/2207.10598} {2207.10598} \BibitemShut {NoStop}%
\bibitem [{\citenamefont {Ekman}(2022)}]{crosscap2}%
  \BibitemOpen
  \bibfield  {author} {\bibinfo {author} {\bibfnamefont {C.}~\bibnamefont {Ekman}},\ }\bibfield  {journal} {\bibinfo  {journal} {arXiv}\ }\href {https://doi.org/10.48550/arXiv.2207.12354} {10.48550/arXiv.2207.12354} (\bibinfo {year} {2022}),\ \Eprint {https://arxiv.org/abs/2207.12354} {2207.12354} \BibitemShut {NoStop}%
\bibitem [{\citenamefont {Udupa}\ \emph {et~al.}(2023)\citenamefont {Udupa}, \citenamefont {Sur}, \citenamefont {Nandy}, \citenamefont {Sen},\ and\ \citenamefont {Sen}}]{Udupa2023Dec}%
  \BibitemOpen
  \bibfield  {author} {\bibinfo {author} {\bibfnamefont {A.}~\bibnamefont {Udupa}}, \bibinfo {author} {\bibfnamefont {S.}~\bibnamefont {Sur}}, \bibinfo {author} {\bibfnamefont {S.}~\bibnamefont {Nandy}}, \bibinfo {author} {\bibfnamefont {A.}~\bibnamefont {Sen}},\ and\ \bibinfo {author} {\bibfnamefont {D.}~\bibnamefont {Sen}},\ }\href {https://doi.org/10.1103/PhysRevB.108.214430} {\bibfield  {journal} {\bibinfo  {journal} {Phys. Rev. B}\ }\textbf {\bibinfo {volume} {108}},\ \bibinfo {pages} {214430} (\bibinfo {year} {2023})}\BibitemShut {NoStop}%
\bibitem [{\citenamefont {ISHIBASHI}(1989)}]{Ishibashi}%
  \BibitemOpen
  \bibfield  {author} {\bibinfo {author} {\bibfnamefont {N.}~\bibnamefont {ISHIBASHI}},\ }\href {https://doi.org/10.1142/S0217732389000320} {\bibfield  {journal} {\bibinfo  {journal} {Modern Physics Letters A}\ }\textbf {\bibinfo {volume} {04}},\ \bibinfo {pages} {251} (\bibinfo {year} {1989})},\ \Eprint {https://arxiv.org/abs/https://doi.org/10.1142/S0217732389000320} {https://doi.org/10.1142/S0217732389000320} \BibitemShut {NoStop}%
\bibitem [{\citenamefont {Majumdar}\ and\ \citenamefont {Ghosh}(1969{\natexlab{a}})}]{MG1}%
  \BibitemOpen
  \bibfield  {author} {\bibinfo {author} {\bibfnamefont {C.~K.}\ \bibnamefont {Majumdar}}\ and\ \bibinfo {author} {\bibfnamefont {D.~K.}\ \bibnamefont {Ghosh}},\ }\href {https://doi.org/10.1063/1.1664978} {\bibfield  {journal} {\bibinfo  {journal} {J. Math. Phys.}\ }\textbf {\bibinfo {volume} {10}},\ \bibinfo {pages} {1388} (\bibinfo {year} {1969}{\natexlab{a}})}\BibitemShut {NoStop}%
\bibitem [{\citenamefont {Majumdar}\ and\ \citenamefont {Ghosh}(1969{\natexlab{b}})}]{MG2}%
  \BibitemOpen
  \bibfield  {author} {\bibinfo {author} {\bibfnamefont {C.~K.}\ \bibnamefont {Majumdar}}\ and\ \bibinfo {author} {\bibfnamefont {D.~K.}\ \bibnamefont {Ghosh}},\ }\href {https://doi.org/10.1063/1.1664979} {\bibfield  {journal} {\bibinfo  {journal} {J. Math. Phys.}\ }\textbf {\bibinfo {volume} {10}},\ \bibinfo {pages} {1399} (\bibinfo {year} {1969}{\natexlab{b}})}\BibitemShut {NoStop}%
\bibitem [{\citenamefont {Anderson}(1973)}]{Anderson1973Feb}%
  \BibitemOpen
  \bibfield  {author} {\bibinfo {author} {\bibfnamefont {P.~W.}\ \bibnamefont {Anderson}},\ }\href {https://doi.org/10.1016/0025-5408(73)90167-0} {\bibfield  {journal} {\bibinfo  {journal} {Mater. Res. Bull.}\ }\textbf {\bibinfo {volume} {8}},\ \bibinfo {pages} {153} (\bibinfo {year} {1973})}\BibitemShut {NoStop}%
\bibitem [{\citenamefont {Anderson}(1987)}]{Anderson1987Mar}%
  \BibitemOpen
  \bibfield  {author} {\bibinfo {author} {\bibfnamefont {P.~W.}\ \bibnamefont {Anderson}},\ }\href {https://doi.org/10.1126/science.235.4793.1196} {\bibfield  {journal} {\bibinfo  {journal} {Science}\ }\textbf {\bibinfo {volume} {235}},\ \bibinfo {pages} {1196} (\bibinfo {year} {1987})}\BibitemShut {NoStop}%
\bibitem [{\citenamefont {Anderson}\ \emph {et~al.}(1987)\citenamefont {Anderson}, \citenamefont {Baskaran}, \citenamefont {Zou},\ and\ \citenamefont {Hsu}}]{Baskaran1987}%
  \BibitemOpen
  \bibfield  {author} {\bibinfo {author} {\bibfnamefont {P.~W.}\ \bibnamefont {Anderson}}, \bibinfo {author} {\bibfnamefont {G.}~\bibnamefont {Baskaran}}, \bibinfo {author} {\bibfnamefont {Z.}~\bibnamefont {Zou}},\ and\ \bibinfo {author} {\bibfnamefont {T.}~\bibnamefont {Hsu}},\ }\href {https://doi.org/10.1103/PhysRevLett.58.2790} {\bibfield  {journal} {\bibinfo  {journal} {Phys. Rev. Lett.}\ }\textbf {\bibinfo {volume} {58}},\ \bibinfo {pages} {2790} (\bibinfo {year} {1987})}\BibitemShut {NoStop}%
\bibitem [{\citenamefont {Baskaran}\ \emph {et~al.}(1993)\citenamefont {Baskaran}, \citenamefont {Zou},\ and\ \citenamefont {Anderson}}]{Baskaran1993Dec}%
  \BibitemOpen
  \bibfield  {author} {\bibinfo {author} {\bibfnamefont {G.}~\bibnamefont {Baskaran}}, \bibinfo {author} {\bibfnamefont {Z.}~\bibnamefont {Zou}},\ and\ \bibinfo {author} {\bibfnamefont {P.~W.}\ \bibnamefont {Anderson}},\ }\href {https://doi.org/10.1016/0038-1098(93)90256-M} {\bibfield  {journal} {\bibinfo  {journal} {Solid State Commun.}\ }\textbf {\bibinfo {volume} {88}},\ \bibinfo {pages} {853} (\bibinfo {year} {1993})}\BibitemShut {NoStop}%
\bibitem [{\citenamefont {Rokhsar}\ and\ \citenamefont {Kivelson}(1988)}]{RK}%
  \BibitemOpen
  \bibfield  {author} {\bibinfo {author} {\bibfnamefont {D.~S.}\ \bibnamefont {Rokhsar}}\ and\ \bibinfo {author} {\bibfnamefont {S.~A.}\ \bibnamefont {Kivelson}},\ }\href {https://doi.org/10.1103/PhysRevLett.61.2376} {\bibfield  {journal} {\bibinfo  {journal} {Phys. Rev. Lett.}\ }\textbf {\bibinfo {volume} {61}},\ \bibinfo {pages} {2376} (\bibinfo {year} {1988})}\BibitemShut {NoStop}%
\bibitem [{\citenamefont {Kivelson}\ \emph {et~al.}(1987)\citenamefont {Kivelson}, \citenamefont {Rokhsar},\ and\ \citenamefont {Sethna}}]{RKS}%
  \BibitemOpen
  \bibfield  {author} {\bibinfo {author} {\bibfnamefont {S.~A.}\ \bibnamefont {Kivelson}}, \bibinfo {author} {\bibfnamefont {D.~S.}\ \bibnamefont {Rokhsar}},\ and\ \bibinfo {author} {\bibfnamefont {J.~P.}\ \bibnamefont {Sethna}},\ }\href {https://doi.org/10.1103/PhysRevB.35.8865} {\bibfield  {journal} {\bibinfo  {journal} {Phys. Rev. B}\ }\textbf {\bibinfo {volume} {35}},\ \bibinfo {pages} {8865} (\bibinfo {year} {1987})}\BibitemShut {NoStop}%
\bibitem [{\citenamefont {Moessner}\ and\ \citenamefont {Sondhi}(2001)}]{RoderichSondhi}%
  \BibitemOpen
  \bibfield  {author} {\bibinfo {author} {\bibfnamefont {R.}~\bibnamefont {Moessner}}\ and\ \bibinfo {author} {\bibfnamefont {S.~L.}\ \bibnamefont {Sondhi}},\ }\href {https://doi.org/10.1103/PhysRevLett.86.1881} {\bibfield  {journal} {\bibinfo  {journal} {Phys. Rev. Lett.}\ }\textbf {\bibinfo {volume} {86}},\ \bibinfo {pages} {1881} (\bibinfo {year} {2001})}\BibitemShut {NoStop}%
\bibitem [{\citenamefont {Balents}(2010)}]{Balents2010Mar}%
  \BibitemOpen
  \bibfield  {author} {\bibinfo {author} {\bibfnamefont {L.}~\bibnamefont {Balents}},\ }\href {https://doi.org/10.1038/nature08917} {\bibfield  {journal} {\bibinfo  {journal} {Nature}\ }\textbf {\bibinfo {volume} {464}},\ \bibinfo {pages} {199} (\bibinfo {year} {2010})}\BibitemShut {NoStop}%
\bibitem [{\citenamefont {Liang}\ \emph {et~al.}(1988)\citenamefont {Liang}, \citenamefont {Doucot},\ and\ \citenamefont {Anderson}}]{tuneRVB}%
  \BibitemOpen
  \bibfield  {author} {\bibinfo {author} {\bibfnamefont {S.}~\bibnamefont {Liang}}, \bibinfo {author} {\bibfnamefont {B.}~\bibnamefont {Doucot}},\ and\ \bibinfo {author} {\bibfnamefont {P.~W.}\ \bibnamefont {Anderson}},\ }\href {https://doi.org/10.1103/PhysRevLett.61.365} {\bibfield  {journal} {\bibinfo  {journal} {Phys. Rev. Lett.}\ }\textbf {\bibinfo {volume} {61}},\ \bibinfo {pages} {365} (\bibinfo {year} {1988})}\BibitemShut {NoStop}%
\bibitem [{\citenamefont {Schecter}\ and\ \citenamefont {Iadecola}(2018)}]{Schecter_Chiral_Index_2018}%
  \BibitemOpen
  \bibfield  {author} {\bibinfo {author} {\bibfnamefont {M.}~\bibnamefont {Schecter}}\ and\ \bibinfo {author} {\bibfnamefont {T.}~\bibnamefont {Iadecola}},\ }\href {https://doi.org/10.1103/PhysRevB.98.035139} {\bibfield  {journal} {\bibinfo  {journal} {Phys. Rev. B}\ }\textbf {\bibinfo {volume} {98}},\ \bibinfo {pages} {035139} (\bibinfo {year} {2018})}\BibitemShut {NoStop}%
\bibitem [{\citenamefont {Karle}\ \emph {et~al.}(2021)\citenamefont {Karle}, \citenamefont {Serbyn},\ and\ \citenamefont {Michailidis}}]{Karle_AreaLaw}%
  \BibitemOpen
  \bibfield  {author} {\bibinfo {author} {\bibfnamefont {V.}~\bibnamefont {Karle}}, \bibinfo {author} {\bibfnamefont {M.}~\bibnamefont {Serbyn}},\ and\ \bibinfo {author} {\bibfnamefont {A.~A.}\ \bibnamefont {Michailidis}},\ }\href {https://doi.org/10.1103/PhysRevLett.127.060602} {\bibfield  {journal} {\bibinfo  {journal} {Phys. Rev. Lett.}\ }\textbf {\bibinfo {volume} {127}},\ \bibinfo {pages} {060602} (\bibinfo {year} {2021})}\BibitemShut {NoStop}%
\bibitem [{\citenamefont {Banerjee}\ and\ \citenamefont {Sen}(2021)}]{senZeromodes}%
  \BibitemOpen
  \bibfield  {author} {\bibinfo {author} {\bibfnamefont {D.}~\bibnamefont {Banerjee}}\ and\ \bibinfo {author} {\bibfnamefont {A.}~\bibnamefont {Sen}},\ }\href {https://doi.org/10.1103/PhysRevLett.126.220601} {\bibfield  {journal} {\bibinfo  {journal} {Phys. Rev. Lett.}\ }\textbf {\bibinfo {volume} {126}},\ \bibinfo {pages} {220601} (\bibinfo {year} {2021})}\BibitemShut {NoStop}%
\bibitem [{\citenamefont {Turner}\ \emph {et~al.}(2024)\citenamefont {Turner}, \citenamefont {Szyniszewski}, \citenamefont {Mukherjee}, \citenamefont {Melendrez}, \citenamefont {Changlani},\ and\ \citenamefont {Pal}}]{Turner2024Jul}%
  \BibitemOpen
  \bibfield  {author} {\bibinfo {author} {\bibfnamefont {C.~J.}\ \bibnamefont {Turner}}, \bibinfo {author} {\bibfnamefont {M.}~\bibnamefont {Szyniszewski}}, \bibinfo {author} {\bibfnamefont {B.}~\bibnamefont {Mukherjee}}, \bibinfo {author} {\bibfnamefont {R.}~\bibnamefont {Melendrez}}, \bibinfo {author} {\bibfnamefont {H.~J.}\ \bibnamefont {Changlani}},\ and\ \bibinfo {author} {\bibfnamefont {A.}~\bibnamefont {Pal}},\ }\bibfield  {journal} {\bibinfo  {journal} {arXiv}\ }\href {https://doi.org/10.48550/arXiv.2407.11956} {10.48550/arXiv.2407.11956} (\bibinfo {year} {2024}),\ \Eprint {https://arxiv.org/abs/2407.11956} {2407.11956} \BibitemShut {NoStop}%
\bibitem [{\citenamefont {Swain}\ \emph {et~al.}(2021)\citenamefont {Swain}, \citenamefont {Tang}, \citenamefont {Foo}, \citenamefont {Khor}, \citenamefont {Lemari{\ifmmode\acute{e}\else\'{e}\fi}}, \citenamefont {Assaad}, \citenamefont {Sengupta},\ and\ \citenamefont {Adam}}]{Swain2021}%
  \BibitemOpen
  \bibfield  {author} {\bibinfo {author} {\bibfnamefont {N.}~\bibnamefont {Swain}}, \bibinfo {author} {\bibfnamefont {H.-K.}\ \bibnamefont {Tang}}, \bibinfo {author} {\bibfnamefont {D.~C.~W.}\ \bibnamefont {Foo}}, \bibinfo {author} {\bibfnamefont {B.~J.~J.}\ \bibnamefont {Khor}}, \bibinfo {author} {\bibfnamefont {G.}~\bibnamefont {Lemari{\ifmmode\acute{e}\else\'{e}\fi}}}, \bibinfo {author} {\bibfnamefont {F.~F.}\ \bibnamefont {Assaad}}, \bibinfo {author} {\bibfnamefont {P.}~\bibnamefont {Sengupta}},\ and\ \bibinfo {author} {\bibfnamefont {S.}~\bibnamefont {Adam}},\ }\bibfield  {journal} {\bibinfo  {journal} {arXiv}\ }\href {https://doi.org/10.48550/arXiv.2106.08587} {10.48550/arXiv.2106.08587} (\bibinfo {year} {2021}),\ \bibinfo {note} {relevant results in v3, published August 2024},\ \Eprint {https://arxiv.org/abs/2106.08587} {2106.08587} \BibitemShut {NoStop}%
\bibitem [{\citenamefont {Melendrez}\ \emph {et~al.}(2025)\citenamefont {Melendrez}, \citenamefont {Mukherjee}, \citenamefont {Szyniszewski}, \citenamefont {Turner}, \citenamefont {Pal},\ and\ \citenamefont {Changlani}}]{Melendrez2025Jan}%
  \BibitemOpen
  \bibfield  {author} {\bibinfo {author} {\bibfnamefont {R.}~\bibnamefont {Melendrez}}, \bibinfo {author} {\bibfnamefont {B.}~\bibnamefont {Mukherjee}}, \bibinfo {author} {\bibfnamefont {M.}~\bibnamefont {Szyniszewski}}, \bibinfo {author} {\bibfnamefont {C.~J.}\ \bibnamefont {Turner}}, \bibinfo {author} {\bibfnamefont {A.}~\bibnamefont {Pal}},\ and\ \bibinfo {author} {\bibfnamefont {H.~J.}\ \bibnamefont {Changlani}},\ }\bibfield  {journal} {\bibinfo  {journal} {arXiv}\ }\href {https://doi.org/10.48550/arXiv.2501.14017} {10.48550/arXiv.2501.14017} (\bibinfo {year} {2025}),\ \Eprint {https://arxiv.org/abs/2501.14017} {2501.14017} \BibitemShut {NoStop}%
\bibitem [{\citenamefont {Hida}(1992{\natexlab{a}})}]{hida_1992}%
  \BibitemOpen
  \bibfield  {author} {\bibinfo {author} {\bibfnamefont {K.}~\bibnamefont {Hida}},\ }\href {https://doi.org/10.1103/PhysRevB.45.2207} {\bibfield  {journal} {\bibinfo  {journal} {Phys. Rev. B}\ }\textbf {\bibinfo {volume} {45}},\ \bibinfo {pages} {2207} (\bibinfo {year} {1992}{\natexlab{a}})}\BibitemShut {NoStop}%
\bibitem [{\citenamefont {Hida}(1992{\natexlab{b}})}]{hida_1992_phasediagram}%
  \BibitemOpen
  \bibfield  {author} {\bibinfo {author} {\bibfnamefont {K.}~\bibnamefont {Hida}},\ }\href {https://doi.org/10.1103/PhysRevB.46.8268} {\bibfield  {journal} {\bibinfo  {journal} {Phys. Rev. B}\ }\textbf {\bibinfo {volume} {46}},\ \bibinfo {pages} {8268} (\bibinfo {year} {1992}{\natexlab{b}})}\BibitemShut {NoStop}%
\bibitem [{\citenamefont {Kennedy}\ and\ \citenamefont {Tasaki}(1992)}]{Kennedy1992}%
  \BibitemOpen
  \bibfield  {author} {\bibinfo {author} {\bibfnamefont {T.}~\bibnamefont {Kennedy}}\ and\ \bibinfo {author} {\bibfnamefont {H.}~\bibnamefont {Tasaki}},\ }\href {https://doi.org/10.1103/PhysRevB.45.304} {\bibfield  {journal} {\bibinfo  {journal} {Phys. Rev. B}\ }\textbf {\bibinfo {volume} {45}},\ \bibinfo {pages} {304} (\bibinfo {year} {1992})}\BibitemShut {NoStop}%
\bibitem [{\citenamefont {Manaka}\ \emph {et~al.}(1997)\citenamefont {Manaka}, \citenamefont {Yamada},\ and\ \citenamefont {Yamaguchi}}]{manaka1997}%
  \BibitemOpen
  \bibfield  {author} {\bibinfo {author} {\bibfnamefont {H.}~\bibnamefont {Manaka}}, \bibinfo {author} {\bibfnamefont {I.}~\bibnamefont {Yamada}},\ and\ \bibinfo {author} {\bibfnamefont {K.}~\bibnamefont {Yamaguchi}},\ }\href {https://doi.org/https://doi.org/10.1143/jpsj.66.564} {\bibfield  {journal} {\bibinfo  {journal} {Journal of the Physical Society of Japan}\ }\textbf {\bibinfo {volume} {66}},\ \bibinfo {pages} {564–567} (\bibinfo {year} {1997})}\BibitemShut {NoStop}%
\bibitem [{Sup()}]{SuppMat}%
  \BibitemOpen
  \href@noop {} {\bibinfo {title} {{See Supplemental Material, which includes details of proofs related to symmetric tensor scars (Appendix A), calculation of R\'enyi-2 entropy (Appendix B) and correlation functions (Appendix C), details on von Neumann entropy (Appendix D), and discussion on the stability against perturbations (Appendix E). Here we additionally refer to Refs.~\cite{Bergholm:2011,Reuvers2018Jun,Lin_perturb}.}}}\BibitemShut {Stop}%
\bibitem [{\citenamefont {Sanada}\ \emph {et~al.}(2023)\citenamefont {Sanada}, \citenamefont {Miao},\ and\ \citenamefont {Katsura}}]{IBS1}%
  \BibitemOpen
  \bibfield  {author} {\bibinfo {author} {\bibfnamefont {K.}~\bibnamefont {Sanada}}, \bibinfo {author} {\bibfnamefont {Y.}~\bibnamefont {Miao}},\ and\ \bibinfo {author} {\bibfnamefont {H.}~\bibnamefont {Katsura}},\ }\href {https://doi.org/10.1103/PhysRevB.108.155102} {\bibfield  {journal} {\bibinfo  {journal} {Phys. Rev. B}\ }\textbf {\bibinfo {volume} {108}},\ \bibinfo {pages} {155102} (\bibinfo {year} {2023})}\BibitemShut {NoStop}%
\bibitem [{\citenamefont {Kazuyuki~Sanada}(2024)}]{IBS2}%
  \BibitemOpen
  \bibfield  {author} {\bibinfo {author} {\bibfnamefont {H.~K.}\ \bibnamefont {Kazuyuki~Sanada}, \bibfnamefont {Yuan~Miao}},\ }\bibfield  {journal} {\bibinfo  {journal} {arXiv}\ }\href {https://doi.org/https://doi.org/10.48550/arXiv.2411.01270} {https://doi.org/10.48550/arXiv.2411.01270} (\bibinfo {year} {2024}),\ \Eprint {https://arxiv.org/abs/2411.01270v1} {2411.01270v1} \BibitemShut {NoStop}%
\bibitem [{\citenamefont {Bao}\ \emph {et~al.}(2020)\citenamefont {Bao}, \citenamefont {Choi},\ and\ \citenamefont {Altman}}]{bao_2020}%
  \BibitemOpen
  \bibfield  {author} {\bibinfo {author} {\bibfnamefont {Y.}~\bibnamefont {Bao}}, \bibinfo {author} {\bibfnamefont {S.}~\bibnamefont {Choi}},\ and\ \bibinfo {author} {\bibfnamefont {E.}~\bibnamefont {Altman}},\ }\href {https://doi.org/10.1103/PhysRevB.101.104301} {\bibfield  {journal} {\bibinfo  {journal} {Phys. Rev. B}\ }\textbf {\bibinfo {volume} {101}},\ \bibinfo {pages} {104301} (\bibinfo {year} {2020})}\BibitemShut {NoStop}%
\bibitem [{\citenamefont {Zhou}\ and\ \citenamefont {Nahum}(2019)}]{zhou_nahum_2018}%
  \BibitemOpen
  \bibfield  {author} {\bibinfo {author} {\bibfnamefont {T.}~\bibnamefont {Zhou}}\ and\ \bibinfo {author} {\bibfnamefont {A.}~\bibnamefont {Nahum}},\ }\href {https://doi.org/10.1103/PhysRevB.99.174205} {\bibfield  {journal} {\bibinfo  {journal} {Phys. Rev. B}\ }\textbf {\bibinfo {volume} {99}},\ \bibinfo {pages} {174205} (\bibinfo {year} {2019})}\BibitemShut {NoStop}%
\bibitem [{\citenamefont {Jian}\ \emph {et~al.}(2020)\citenamefont {Jian}, \citenamefont {You}, \citenamefont {Vasseur},\ and\ \citenamefont {Ludwig}}]{jian_2020}%
  \BibitemOpen
  \bibfield  {author} {\bibinfo {author} {\bibfnamefont {C.-M.}\ \bibnamefont {Jian}}, \bibinfo {author} {\bibfnamefont {Y.-Z.}\ \bibnamefont {You}}, \bibinfo {author} {\bibfnamefont {R.}~\bibnamefont {Vasseur}},\ and\ \bibinfo {author} {\bibfnamefont {A.~W.~W.}\ \bibnamefont {Ludwig}},\ }\href {https://doi.org/10.1103/PhysRevB.101.104302} {\bibfield  {journal} {\bibinfo  {journal} {Phys. Rev. B}\ }\textbf {\bibinfo {volume} {101}},\ \bibinfo {pages} {104302} (\bibinfo {year} {2020})}\BibitemShut {NoStop}%
\bibitem [{\citenamefont {Alet}\ and\ \citenamefont {Laflorencie}(2018)}]{MBL1}%
  \BibitemOpen
  \bibfield  {author} {\bibinfo {author} {\bibfnamefont {F.}~\bibnamefont {Alet}}\ and\ \bibinfo {author} {\bibfnamefont {N.}~\bibnamefont {Laflorencie}},\ }\href {https://doi.org/10.1016/j.crhy.2018.03.003} {\bibfield  {journal} {\bibinfo  {journal} {C. R. Phys.}\ }\textbf {\bibinfo {volume} {19}},\ \bibinfo {pages} {498} (\bibinfo {year} {2018})}\BibitemShut {NoStop}%
\bibitem [{\citenamefont {Vasseur}\ \emph {et~al.}(2019)\citenamefont {Vasseur}, \citenamefont {Potter}, \citenamefont {You},\ and\ \citenamefont {Ludwig}}]{ET1}%
  \BibitemOpen
  \bibfield  {author} {\bibinfo {author} {\bibfnamefont {R.}~\bibnamefont {Vasseur}}, \bibinfo {author} {\bibfnamefont {A.~C.}\ \bibnamefont {Potter}}, \bibinfo {author} {\bibfnamefont {Y.-Z.}\ \bibnamefont {You}},\ and\ \bibinfo {author} {\bibfnamefont {A.~W.~W.}\ \bibnamefont {Ludwig}},\ }\href {https://doi.org/10.1103/PhysRevB.100.134203} {\bibfield  {journal} {\bibinfo  {journal} {Phys. Rev. B}\ }\textbf {\bibinfo {volume} {100}},\ \bibinfo {pages} {134203} (\bibinfo {year} {2019})}\BibitemShut {NoStop}%
\bibitem [{\citenamefont {Skinner}\ \emph {et~al.}(2019)\citenamefont {Skinner}, \citenamefont {Ruhman},\ and\ \citenamefont {Nahum}}]{ET2}%
  \BibitemOpen
  \bibfield  {author} {\bibinfo {author} {\bibfnamefont {B.}~\bibnamefont {Skinner}}, \bibinfo {author} {\bibfnamefont {J.}~\bibnamefont {Ruhman}},\ and\ \bibinfo {author} {\bibfnamefont {A.}~\bibnamefont {Nahum}},\ }\href {https://doi.org/10.1103/PhysRevX.9.031009} {\bibfield  {journal} {\bibinfo  {journal} {Phys. Rev. X}\ }\textbf {\bibinfo {volume} {9}},\ \bibinfo {pages} {031009} (\bibinfo {year} {2019})}\BibitemShut {NoStop}%
\bibitem [{\citenamefont {Li}\ \emph {et~al.}(2018)\citenamefont {Li}, \citenamefont {Chen},\ and\ \citenamefont {Fisher}}]{ET3}%
  \BibitemOpen
  \bibfield  {author} {\bibinfo {author} {\bibfnamefont {Y.}~\bibnamefont {Li}}, \bibinfo {author} {\bibfnamefont {X.}~\bibnamefont {Chen}},\ and\ \bibinfo {author} {\bibfnamefont {M.~P.~A.}\ \bibnamefont {Fisher}},\ }\href {https://doi.org/10.1103/PhysRevB.98.205136} {\bibfield  {journal} {\bibinfo  {journal} {Phys. Rev. B}\ }\textbf {\bibinfo {volume} {98}},\ \bibinfo {pages} {205136} (\bibinfo {year} {2018})}\BibitemShut {NoStop}%
\bibitem [{\citenamefont {Chan}\ \emph {et~al.}(2019)\citenamefont {Chan}, \citenamefont {Nandkishore}, \citenamefont {Pretko},\ and\ \citenamefont {Smith}}]{ET4}%
  \BibitemOpen
  \bibfield  {author} {\bibinfo {author} {\bibfnamefont {A.}~\bibnamefont {Chan}}, \bibinfo {author} {\bibfnamefont {R.~M.}\ \bibnamefont {Nandkishore}}, \bibinfo {author} {\bibfnamefont {M.}~\bibnamefont {Pretko}},\ and\ \bibinfo {author} {\bibfnamefont {G.}~\bibnamefont {Smith}},\ }\href {https://doi.org/10.1103/PhysRevB.99.224307} {\bibfield  {journal} {\bibinfo  {journal} {Phys. Rev. B}\ }\textbf {\bibinfo {volume} {99}},\ \bibinfo {pages} {224307} (\bibinfo {year} {2019})}\BibitemShut {NoStop}%
\bibitem [{\citenamefont {Szyniszewski}\ \emph {et~al.}(2019)\citenamefont {Szyniszewski}, \citenamefont {Romito},\ and\ \citenamefont {Schomerus}}]{ET5}%
  \BibitemOpen
  \bibfield  {author} {\bibinfo {author} {\bibfnamefont {M.}~\bibnamefont {Szyniszewski}}, \bibinfo {author} {\bibfnamefont {A.}~\bibnamefont {Romito}},\ and\ \bibinfo {author} {\bibfnamefont {H.}~\bibnamefont {Schomerus}},\ }\href {https://doi.org/10.1103/PhysRevB.100.064204} {\bibfield  {journal} {\bibinfo  {journal} {Phys. Rev. B}\ }\textbf {\bibinfo {volume} {100}},\ \bibinfo {pages} {064204} (\bibinfo {year} {2019})}\BibitemShut {NoStop}%
\bibitem [{\citenamefont {Chitambar}\ \emph {et~al.}(2014)\citenamefont {Chitambar}, \citenamefont {Leung}, \citenamefont {Man{\ifmmode\check{c}\else\v{c}\fi}inska}, \citenamefont {Ozols},\ and\ \citenamefont {Winter}}]{Chitambar2014}%
  \BibitemOpen
  \bibfield  {author} {\bibinfo {author} {\bibfnamefont {E.}~\bibnamefont {Chitambar}}, \bibinfo {author} {\bibfnamefont {D.}~\bibnamefont {Leung}}, \bibinfo {author} {\bibfnamefont {L.}~\bibnamefont {Man{\ifmmode\check{c}\else\v{c}\fi}inska}}, \bibinfo {author} {\bibfnamefont {M.}~\bibnamefont {Ozols}},\ and\ \bibinfo {author} {\bibfnamefont {A.}~\bibnamefont {Winter}},\ }\href {https://doi.org/10.1007/s00220-014-1953-9} {\bibfield  {journal} {\bibinfo  {journal} {Commun. Math. Phys.}\ }\textbf {\bibinfo {volume} {328}},\ \bibinfo {pages} {303} (\bibinfo {year} {2014})}\BibitemShut {NoStop}%
\bibitem [{\citenamefont {Huse}\ \emph {et~al.}(2013)\citenamefont {Huse}, \citenamefont {Nandkishore}, \citenamefont {Oganesyan}, \citenamefont {Pal},\ and\ \citenamefont {Sondhi}}]{Huse2013Jul}%
  \BibitemOpen
  \bibfield  {author} {\bibinfo {author} {\bibfnamefont {D.~A.}\ \bibnamefont {Huse}}, \bibinfo {author} {\bibfnamefont {R.}~\bibnamefont {Nandkishore}}, \bibinfo {author} {\bibfnamefont {V.}~\bibnamefont {Oganesyan}}, \bibinfo {author} {\bibfnamefont {A.}~\bibnamefont {Pal}},\ and\ \bibinfo {author} {\bibfnamefont {S.~L.}\ \bibnamefont {Sondhi}},\ }\href {https://doi.org/10.1103/PhysRevB.88.014206} {\bibfield  {journal} {\bibinfo  {journal} {Phys. Rev. B}\ }\textbf {\bibinfo {volume} {88}},\ \bibinfo {pages} {014206} (\bibinfo {year} {2013})}\BibitemShut {NoStop}%
\bibitem [{\citenamefont {Iadecola}\ \emph {et~al.}(2019)\citenamefont {Iadecola}, \citenamefont {Schecter},\ and\ \citenamefont {Xu}}]{MagnonScar}%
  \BibitemOpen
  \bibfield  {author} {\bibinfo {author} {\bibfnamefont {T.}~\bibnamefont {Iadecola}}, \bibinfo {author} {\bibfnamefont {M.}~\bibnamefont {Schecter}},\ and\ \bibinfo {author} {\bibfnamefont {S.}~\bibnamefont {Xu}},\ }\href {https://doi.org/10.1103/PhysRevB.100.184312} {\bibfield  {journal} {\bibinfo  {journal} {Phys. Rev. B}\ }\textbf {\bibinfo {volume} {100}},\ \bibinfo {pages} {184312} (\bibinfo {year} {2019})}\BibitemShut {NoStop}%
\bibitem [{\citenamefont {Desaules}\ \emph {et~al.}(2022)\citenamefont {Desaules}, \citenamefont {Pietracaprina}, \citenamefont {Papi{\ifmmode\acute{c}\else\'{c}\fi}}, \citenamefont {Goold},\ and\ \citenamefont {Pappalardi}}]{Desaules2022Jul}%
  \BibitemOpen
  \bibfield  {author} {\bibinfo {author} {\bibfnamefont {J.-Y.}\ \bibnamefont {Desaules}}, \bibinfo {author} {\bibfnamefont {F.}~\bibnamefont {Pietracaprina}}, \bibinfo {author} {\bibfnamefont {Z.}~\bibnamefont {Papi{\ifmmode\acute{c}\else\'{c}\fi}}}, \bibinfo {author} {\bibfnamefont {J.}~\bibnamefont {Goold}},\ and\ \bibinfo {author} {\bibfnamefont {S.}~\bibnamefont {Pappalardi}},\ }\href {https://doi.org/10.1103/PhysRevLett.129.020601} {\bibfield  {journal} {\bibinfo  {journal} {Phys. Rev. Lett.}\ }\textbf {\bibinfo {volume} {129}},\ \bibinfo {pages} {020601} (\bibinfo {year} {2022})}\BibitemShut {NoStop}%
\bibitem [{\citenamefont {Gotta}\ \emph {et~al.}(2023)\citenamefont {Gotta}, \citenamefont {Moudgalya},\ and\ \citenamefont {Mazza}}]{asymptotic}%
  \BibitemOpen
  \bibfield  {author} {\bibinfo {author} {\bibfnamefont {L.}~\bibnamefont {Gotta}}, \bibinfo {author} {\bibfnamefont {S.}~\bibnamefont {Moudgalya}},\ and\ \bibinfo {author} {\bibfnamefont {L.}~\bibnamefont {Mazza}},\ }\href {https://doi.org/10.1103/PhysRevLett.131.190401} {\bibfield  {journal} {\bibinfo  {journal} {Phys. Rev. Lett.}\ }\textbf {\bibinfo {volume} {131}},\ \bibinfo {pages} {190401} (\bibinfo {year} {2023})}\BibitemShut {NoStop}%
\bibitem [{\citenamefont {Ho}\ \emph {et~al.}(2019)\citenamefont {Ho}, \citenamefont {Choi}, \citenamefont {Pichler},\ and\ \citenamefont {Lukin}}]{Ho:2019}%
  \BibitemOpen
  \bibfield  {author} {\bibinfo {author} {\bibfnamefont {W.~W.}\ \bibnamefont {Ho}}, \bibinfo {author} {\bibfnamefont {S.}~\bibnamefont {Choi}}, \bibinfo {author} {\bibfnamefont {H.}~\bibnamefont {Pichler}},\ and\ \bibinfo {author} {\bibfnamefont {M.~D.}\ \bibnamefont {Lukin}},\ }\href {https://doi.org/10.1103/physrevlett.122.040603} {\bibfield  {journal} {\bibinfo  {journal} {Phys. Rev. Lett.}\ }\textbf {\bibinfo {volume} {122}},\ \bibinfo {pages} {040603} (\bibinfo {year} {2019})}\BibitemShut {NoStop}%
\bibitem [{\citenamefont {Turner}\ \emph {et~al.}(2021)\citenamefont {Turner}, \citenamefont {Desaules}, \citenamefont {Bull},\ and\ \citenamefont {Papić}}]{Turner:2021}%
  \BibitemOpen
  \bibfield  {author} {\bibinfo {author} {\bibfnamefont {C.~J.}\ \bibnamefont {Turner}}, \bibinfo {author} {\bibfnamefont {J.-Y.}\ \bibnamefont {Desaules}}, \bibinfo {author} {\bibfnamefont {K.}~\bibnamefont {Bull}},\ and\ \bibinfo {author} {\bibfnamefont {Z.}~\bibnamefont {Papić}},\ }\href {https://doi.org/10.1103/physrevx.11.021021} {\bibfield  {journal} {\bibinfo  {journal} {Phys. Rev. X}\ }\textbf {\bibinfo {volume} {11}},\ \bibinfo {pages} {021021} (\bibinfo {year} {2021})}\BibitemShut {NoStop}%
\bibitem [{\citenamefont {Chen}\ \emph {et~al.}(2022)\citenamefont {Chen}, \citenamefont {Burdick}, \citenamefont {Yao}, \citenamefont {Orth},\ and\ \citenamefont {Iadecola}}]{Chen2022Oct}%
  \BibitemOpen
  \bibfield  {author} {\bibinfo {author} {\bibfnamefont {I.-C.}\ \bibnamefont {Chen}}, \bibinfo {author} {\bibfnamefont {B.}~\bibnamefont {Burdick}}, \bibinfo {author} {\bibfnamefont {Y.}~\bibnamefont {Yao}}, \bibinfo {author} {\bibfnamefont {P.~P.}\ \bibnamefont {Orth}},\ and\ \bibinfo {author} {\bibfnamefont {T.}~\bibnamefont {Iadecola}},\ }\href {https://doi.org/10.1103/PhysRevResearch.4.043027} {\bibfield  {journal} {\bibinfo  {journal} {Phys. Rev. Res.}\ }\textbf {\bibinfo {volume} {4}},\ \bibinfo {pages} {043027} (\bibinfo {year} {2022})}\BibitemShut {NoStop}%
\bibitem [{\citenamefont {Gustafson}\ \emph {et~al.}(2023)\citenamefont {Gustafson}, \citenamefont {Li}, \citenamefont {Khan}, \citenamefont {Kim}, \citenamefont {Kurkcuoglu}, \citenamefont {Alam}, \citenamefont {Orth}, \citenamefont {Rahmani},\ and\ \citenamefont {Iadecola}}]{Gustafson2023Nov}%
  \BibitemOpen
  \bibfield  {author} {\bibinfo {author} {\bibfnamefont {E.~J.}\ \bibnamefont {Gustafson}}, \bibinfo {author} {\bibfnamefont {A.~C.~Y.}\ \bibnamefont {Li}}, \bibinfo {author} {\bibfnamefont {A.}~\bibnamefont {Khan}}, \bibinfo {author} {\bibfnamefont {J.}~\bibnamefont {Kim}}, \bibinfo {author} {\bibfnamefont {D.~M.}\ \bibnamefont {Kurkcuoglu}}, \bibinfo {author} {\bibfnamefont {M.~S.}\ \bibnamefont {Alam}}, \bibinfo {author} {\bibfnamefont {P.~P.}\ \bibnamefont {Orth}}, \bibinfo {author} {\bibfnamefont {A.}~\bibnamefont {Rahmani}},\ and\ \bibinfo {author} {\bibfnamefont {T.}~\bibnamefont {Iadecola}},\ }\href {https://doi.org/10.22331/q-2023-11-07-1171} {\bibfield  {journal} {\bibinfo  {journal} {Quantum}\ }\textbf {\bibinfo {volume} {7}},\ \bibinfo {pages} {1171} (\bibinfo {year} {2023})}\BibitemShut {NoStop}%
\bibitem [{\citenamefont {Dong}\ \emph {et~al.}(2023)\citenamefont {Dong}, \citenamefont {Desaules}, \citenamefont {Gao}, \citenamefont {Wang}, \citenamefont {Guo}, \citenamefont {Chen}, \citenamefont {Zou}, \citenamefont {Jin}, \citenamefont {Zhu}, \citenamefont {Zhang}, \citenamefont {Li}, \citenamefont {Wang}, \citenamefont {Guo}, \citenamefont {Zhang}, \citenamefont {Ying},\ and\ \citenamefont {Papi{\ifmmode\acute{c}\else\'{c}\fi}}}]{Dong2023Dec}%
  \BibitemOpen
  \bibfield  {author} {\bibinfo {author} {\bibfnamefont {H.}~\bibnamefont {Dong}}, \bibinfo {author} {\bibfnamefont {J.-Y.}\ \bibnamefont {Desaules}}, \bibinfo {author} {\bibfnamefont {Y.}~\bibnamefont {Gao}}, \bibinfo {author} {\bibfnamefont {N.}~\bibnamefont {Wang}}, \bibinfo {author} {\bibfnamefont {Z.}~\bibnamefont {Guo}}, \bibinfo {author} {\bibfnamefont {J.}~\bibnamefont {Chen}}, \bibinfo {author} {\bibfnamefont {Y.}~\bibnamefont {Zou}}, \bibinfo {author} {\bibfnamefont {F.}~\bibnamefont {Jin}}, \bibinfo {author} {\bibfnamefont {X.}~\bibnamefont {Zhu}}, \bibinfo {author} {\bibfnamefont {P.}~\bibnamefont {Zhang}}, \bibinfo {author} {\bibfnamefont {H.}~\bibnamefont {Li}}, \bibinfo {author} {\bibfnamefont {Z.}~\bibnamefont {Wang}}, \bibinfo {author} {\bibfnamefont {Q.}~\bibnamefont {Guo}}, \bibinfo {author} {\bibfnamefont {J.}~\bibnamefont {Zhang}}, \bibinfo {author} {\bibfnamefont {L.}~\bibnamefont {Ying}},\ and\ \bibinfo {author} {\bibfnamefont {Z.}~\bibnamefont {Papi{\ifmmode\acute{c}\else\'{c}\fi}}},\
  }\bibfield  {journal} {\bibinfo  {journal} {Sci. Adv.}\ }\textbf {\bibinfo {volume} {9}},\ \href {https://doi.org/10.1126/sciadv.adj3822} {10.1126/sciadv.adj3822} (\bibinfo {year} {2023})\BibitemShut {NoStop}%
\bibitem [{\citenamefont {Mukherjee}\ \emph {et~al.}(2026)\citenamefont {Mukherjee}, \citenamefont {Turner}, \citenamefont {Szyniszewski},\ and\ \citenamefont {Pal}}]{Data}%
  \BibitemOpen
  \bibfield  {author} {\bibinfo {author} {\bibfnamefont {B.}~\bibnamefont {Mukherjee}}, \bibinfo {author} {\bibfnamefont {C.~J.}\ \bibnamefont {Turner}}, \bibinfo {author} {\bibfnamefont {M.}~\bibnamefont {Szyniszewski}},\ and\ \bibinfo {author} {\bibfnamefont {A.}~\bibnamefont {Pal}},\ }\href {https://doi.org/10.25446/oxford.31324879} {10.25446/oxford.31324879} (\bibinfo {year} {2026})\BibitemShut {NoStop}%
\bibitem [{\citenamefont {Bergholm}\ and\ \citenamefont {Biamonte}(2011)}]{Bergholm:2011}%
  \BibitemOpen
  \bibfield  {author} {\bibinfo {author} {\bibfnamefont {V.}~\bibnamefont {Bergholm}}\ and\ \bibinfo {author} {\bibfnamefont {J.~D.}\ \bibnamefont {Biamonte}},\ }\href {https://doi.org/10.1088/1751-8113/44/24/245304} {\bibfield  {journal} {\bibinfo  {journal} {J. Phys. A: Math. Theor.}\ }\textbf {\bibinfo {volume} {44}},\ \bibinfo {pages} {245304} (\bibinfo {year} {2011})}\BibitemShut {NoStop}%
\bibitem [{\citenamefont {Reuvers}(2018)}]{Reuvers2018Jun}%
  \BibitemOpen
  \bibfield  {author} {\bibinfo {author} {\bibfnamefont {R.}~\bibnamefont {Reuvers}},\ }\href {https://doi.org/10.1098/rspa.2018.0023} {\bibfield  {journal} {\bibinfo  {journal} {Proc. R. Soc. A.}\ }\textbf {\bibinfo {volume} {474}},\ \bibinfo {pages} {20180023} (\bibinfo {year} {2018})}\BibitemShut {NoStop}%
\bibitem [{\citenamefont {Lin}\ \emph {et~al.}(2020)\citenamefont {Lin}, \citenamefont {Chandran},\ and\ \citenamefont {Motrunich}}]{Lin_perturb}%
  \BibitemOpen
  \bibfield  {author} {\bibinfo {author} {\bibfnamefont {C.-J.}\ \bibnamefont {Lin}}, \bibinfo {author} {\bibfnamefont {A.}~\bibnamefont {Chandran}},\ and\ \bibinfo {author} {\bibfnamefont {O.~I.}\ \bibnamefont {Motrunich}},\ }\href {https://doi.org/10.1103/PhysRevResearch.2.033044} {\bibfield  {journal} {\bibinfo  {journal} {Phys. Rev. Res.}\ }\textbf {\bibinfo {volume} {2}},\ \bibinfo {pages} {033044} (\bibinfo {year} {2020})}\BibitemShut {NoStop}%
\end{thebibliography}%

\clearpage


\appendix

\setcounter{page}{1}

\title{Supplemental material for ``Symmetric tensor scars with tunable entanglement from volume to area law"}

\maketitle

\renewcommand\thefigure{S\arabic{figure}}
\setcounter{figure}{0}
\setcounter{equation}{0}
\renewcommand{\thetable}{S.\Roman{table}}
\setcounter{table}{0}

\setcounter{section}{0}
\setcounter{secnumdepth}{2}
\renewcommand{\thesection}{\Alph{section}}
\setcounter{subsection}{0}
\renewcommand{\thesubsection}{\Roman{subsection}}

\titleformat{\section}[block]{\normalfont\bfseries\filcenter}{Appendix \thesection:}{.5em}{}[]
\titlespacing{\section}{0ex}{*4}{*2.5}

\titleformat{\subsection}[block]{\normalfont\bfseries\filcenter}{\thesection.\thesubsection:}{.5em}{}[]
\titlespacing{\subsection}{0ex}{*2}{*1}

\section{Symmetric-tensor scar eigenstates}
\label{app:proof}

\subsection{Root states are zero-energy eigenstates}

Recall that we can turn a state $v$ of two spin-1/2 degrees of freedom into a state in the tensor power, which we refer to as a \emph{root state}, using a multilinear map known as the symmetric tensor power,
\begin{align}
  v \mapsto \ket{\Psi(v)} = v^{\otimes N}
\end{align}
where we interpret the first spin-$\frac{1}{2}$ factor (i.e.\@ the left half of the first copy of $v$) as site $1$ of the chain and the second factor as site $1 {+} N$, so each factor of $v$ connects two sites on opposite sides of the system.
There is some arbitrariness in the assignment of sites as being either left or right factors, however, eventually, we will restrict to $v$ where this assignment ultimately only amounts to an overall phase.

In this subsection, we will show that if $v$ is either a singlet state or a triplet state -- but not a superposition of the two -- and provided the half-system size $N$ is odd, then the root state $\ket{\Psi(v)}$ is an exact zero-energy eigenstate.
To see this we first write the Hamiltonian as an alternating sum of swaps.
This can be done because the identity and the swap operator for a complete basis for the SU(2)-invariant operators of $\tfrac{1}{2} \otimes \tfrac{1}{2}$ and the identity component vanishes since there are exactly as many positive terms as there are negative terms.
Each swap can then be paired with an antipodal swap on the opposite side of the system which has an opposite sign due to the restriction to odd $N$.

We understand the action of $H$ by means of a diagrammatic interpretation of $\ket{\Psi(v)}$ similar to the categorical quantum circuits~\cite{Bergholm:2011}.
Each site of the system is a point that is connected to the antipodal point through an oriented strand representing the state $v$.
The orientation records which end of the strand is the left spin-$\frac{1}{2}$ factor of $v$ and which is the right.
A swap term in the Hamiltonian acts by swapping the strand connectivity of the points, creating a diagram where those two strands have become `uncrossed'.
\begin{align}
  \mathrm{SWAP}_{i,j}\!\left[
\begin{tikzpicture}[baseline=0,>={Stealth[scale=1]},scale=0.7]
  \draw[color=gray] (0,0) circle [radius=1];
  \draw[color=black] ({cos(-20)},{sin(-20)}) node (D) {\textbullet};
  \draw[color=black] ({cos(-20)},{sin(-20)}) node[anchor=north west] {\small $j{+}N$};
  \draw[color=black] ({-cos(20)},{sin(20)}) node (A) {\textbullet};
  \draw[color=black] ({-cos(20)},{sin(20)}) node[anchor=south east] {\small $j$};
  \begin{scope}[thick,decoration={markings,mark=at position 0.35 with {\arrow{>}}}]
    \draw[postaction={decorate}] (A.center) -- (D.center);
    \draw[preaction={draw, line width=4pt, white},postaction={decorate}] ({cos(20)},{sin(20)}) -- ({-cos(-20)},{sin(-20)}) ;
  \end{scope}
  \draw[color=black] ({-cos(-20)},{sin(-20)}) node (B) {\textbullet};
  \draw[color=black] ({-cos(-20)},{sin(-20)}) node[anchor=north east] {\small $i$};
  \draw[color=black] ({cos(20)},{sin(20)}) node (C) {\textbullet};
  \draw[color=black] ({cos(20)},{sin(20)}) node[anchor=south west] {\small $i{+}N$};
\end{tikzpicture}\right]
{=}
\begin{tikzpicture}[baseline=0,>={Stealth[scale=1]},scale=0.7]
  \draw[color=gray] (0,0) circle [radius=1];
  \draw[color=black] ({cos(20)},{sin(20)}) node (C) {\textbullet};
  \draw[color=black] ({cos(20)},{sin(20)}) node[anchor=south west] {\small $i{+}N$};
  \draw[color=black] ({cos(-20)},{sin(-20)}) node (D) {\textbullet};
  \draw[color=black] ({cos(-20)},{sin(-20)}) node[anchor=north west] {\small $j{+}N$};
  \draw[color=black] ({-cos(20)},{sin(20)}) node (A) {\textbullet};
  \draw[color=black] ({-cos(20)},{sin(20)}) node[anchor=south east] {\small $j$};
  \draw[color=black] ({-cos(-20)},{sin(-20)}) node (B) {\textbullet};
  \draw[color=black] ({-cos(-20)},{sin(-20)}) node[anchor=north east] {\small $i$};
  \begin{scope}[thick,decoration={markings,mark=at position 0.5 with {\arrow{>}}}]
    \draw[postaction={decorate}] (B.center) -- (D.center);
    \draw[postaction={decorate}] (C.center) -- (A.center);
  \end{scope}
\end{tikzpicture}
\label{eq:swap_strands}
\end{align}
If both strands affected by a swap are orientated either both towards or both away from the swap gate then the diagram from the antipodal swap is identical and the opposite sign in the Hamiltonian causes the contributions to cancel out.
If the strands have instead opposite orientations [as in \Cref{eq:swap_strands}], then the two diagrams are related by reversing both of the orientations.
If we interpret $v$ as a matrix we can write a necessary and sufficient condition for the diagrams to cancel out as
$
  v^T \otimes v - v \otimes v^T = 0
$.
If $v$ is in the singlet representation then $v = -v^T$ or if $v$ is in the triplet representation then $v = v^T$ and in either case the condition is satisfied.
However, any superposition of these two possibilities will fail to create a solution.

If the range of Hamiltonian terms $r$ is odd (we focus on the case $r=1$) then it is not possible to arrange the strands such that the orientations are always either both towards or both away from each swap, therefore this becomes a restriction on the eigenstates.
For even separations $r$ however this is possible, and consequently, it is possible to take linear combinations across the singlet and triplet representations and still obtain a zero-energy eigenstate provided you choose an appropriate orientation.
Additionally, for even separations, you can choose to put different states on the even and odd factors of the tensor-product which further still expands the space of exact eigenstates.

\subsection{Symmetric tensor scars and permutation-invariant bases}
We will show that the span of the root states is the vector space consisting of all symmetric tensors over spin-1 ($V=V_{S=1}$), which is denoted $\Sym^{N}\left(V\right)$ and has dimension,
\begin{align}
  \dim\Sym^{N}\left(V_{S=1}\right) = \binom{N+2}{2}\text{.}
\end{align}
Obviously, $\Span(v^{\otimes N}) \le \Sym^N V$, because each root state is invariant under the permutation action.
Later, we will establish the reverse inequality and therefore equality.
Let $d = \dim V$ and $\{\ket{T_k}\}_{k=1,\ldots,d}$ be a complete linearly independent basis for $V$.
Recall that using this basis we can construct a basis of permutation-invariant states,
\begin{align}
  \ket{T_{[i]}} =
  \frac{1}{\mathcal{N}_c} \sum_{\sigma \in S} \sigma\big(\ket{T_1}^{\otimes n_1} \otimes \cdots \otimes \ket{T_d}^{\otimes n_d}\big)
  \text{,}
  \label{eq:supp_permutation_basis}
\end{align}
We will also show that this basis is complete for $\Sym(V)$.

For any symmetric tensor state $\ket{\phi}$ we can define an associated polynomial $f_\phi : \mathbb{C}^d \rightarrow \mathbb{C}$ by $f_\phi(z) = \langle\bar{z}^{\otimes N} {\mid} \phi\rangle$ in indeterminates $z=(z_1,\ldots,z_d)$ where $\bra{\bar{z}} = \sum_{k=1} z_k \bra{T_k}$.
That this is a degree-$N$ homogeneous polynomial as can be seen by expanding $\ket{\phi}$ in the product basis.
For examples of these associated polynomials, if $\phi$ is a root state then $f_\phi$ is a power of a linear form, and if $\phi$ comes from the permutation-symmetric basis of \Cref{eq:supp_permutation_basis} then $f_\phi$ is a monomial.
We can also turn any of these polynomials back into a symmetric tensor state by substituting for each monomial the unique corresponding permutation-invariant basis state.
This establishes a linear isomorphism between the space of degree-$N$ homogeneous polynomials and the symmetric tensor space.

Clearly, the degree-$N$ monomials form a complete linearly independent basis for the degree-$N$ polynomials.
Hence, through the isomorphism, the permutation basis is also complete for the symmetric tensors.

We can also understand the relationship between the root states and $\Sym(V)$ using the associated polynomials.
For a root state $v$, the associated polynomial $f_{v^{\otimes N}}(z)$ can also be viewed dually as a polynomial in the components of $v$ in the basis for $V$,
\begin{align}
  \tilde{f}_z(v) = f_{v^{\otimes N}}(z) =\bigg(\sum_{k=1}^d z_k v_k\bigg)^N
  \text{.}
\end{align}
The coefficients of $\tilde{f}_z(v)$ can then be found in two ways, first by using the binomial theorem (\Cref{eq:poly_binomial}) and second by the use of the Cauchy integral formula (\Cref{eq:poly_cauchy}),
\begin{align}
  [v_1^{n_1}\cdots v_d^{n_d}] \tilde{f}_z(v)
  &= \frac{N!}{n_1!\cdots n_d!} z_1^{n_1} \cdots z_d^{n_d} \label{eq:poly_binomial}\\
  &= \frac{1}{(2\pi i)^d}\oint_{\partial D} \frac{\tilde{f}_z(v)\,\,\mathrm{d}^d v}{v_1^{n_1+1} \cdots v_d^{n_d+1}}  \label{eq:poly_cauchy}
  \text{,}
\end{align}
where $D$ is a polydisk enclosing the origin in the standard manner.
Despite our use of complex analysis, the result here is really one of algebraic geometry and is a general statement concerning polynomial rings, but we consider complex analysis a more widely familiar tool.
Hence, the monomials (\Cref{eq:poly_binomial}) are linear combinations of powers of linear forms (\Cref{eq:poly_cauchy}), and therefore, by using the isomorphism, every symmetric tensor state is in the linear span of the root states.
Since the root states are zero-energy eigenvectors of the Hamiltonian, we may conclude that every symmetric tensor state is in fact also a zero-energy eigenvector -- including, for example, any basis vector following the permutation-invariant construction.

We find it interesting to comment that the relationship between the permutation-invariant basis and the frame of root states mirrors the relationship between the mean-field TDVP frame~\cite{Ho:2019} used to study scar states in the PXP model and the permutation-invariant scar quasimodes~\cite{Turner:2021} which result from the quantization of that semiclassical frame.

\section{Calculation of \texorpdfstring{$S^{(2)}$}{S(2)} with polynomial resources and asymptotic analysis}
\label{app:R2}

In this section, we show how the second R\'enyi entropy can be calculated in a computationally efficient manner for both the Bell and conventional bases, although the general procedure is not really specific to those bases.
It is frequently found 
that (at least low order) R\'enyi entropies are significantly easier to obtain than von Neumann entropies even to the point where the best means of obtaining the von Neumann entropy involves calculating all infinitely many R\'enyi entropies before analytic continuation.
We will see that the possibility of polynomial-time evaluation of these entropies ultimately comes from the space of states being isomorphic to a certain operator algebra with a polynomially-sized dimension and the coefficients of the multiplication map being easy to calculate.

We first reinterpret the singlet and triplet states as operators, by use of an analog to the Choi-Jamiolkowski isomorphism~\cite{Bergholm:2011}, 
\begin{align}
    &&\ket{T_X^B}&\mapsto\frac{\sigma_X}{\sqrt{2}}, \quad \ket{T_{+}^C}\mapsto\sigma_{+},\nonumber\\
    \ket{S}&\mapsto\frac{\mathbb{I}}{\sqrt{2}},&\ket{T_Y^B}&\mapsto\frac{\sigma_Y}{\sqrt{2}}, \quad \ket{T_{-}^C}\mapsto\sigma_{-},\nonumber\\
    &&\ket{T_Z^B}&\mapsto\frac{\sigma_Z}{\sqrt{2}}, \quad \ket{T_Z^C}\mapsto\frac{\sigma_Z}{\sqrt{2}}
    \text{.}
    \label{eq:reps}
\end{align}
This operation is similar to the \emph{reshaping} of a vector into a matrix -- however, unlike reshaping, it is a basis-independent operation and reveals the symmetry of the resulting algebra.

We have, $S^{(2)}=-\ln{\Tr[\rho^2]}$ where $\rho$ is the reduced density matrix of the subsystem $A$, given by
\begin{equation}
  \rho = \Tr_B (\ket{\Psi} \bra{\Psi}) = \Psi \Psi^\dag.
\end{equation}
Here $B$ represents the environment, $\ket{\Psi}$ and $\Psi$ are respectively the symmetric tensor state and the associated linear operator representation of it.
The subsystem $A$ is assumed to be precisely one half of the system as if we were to cut it into two equal intervals $A$ and $B$, each of length $N$.
This enables the straightforward use of the isomorphism because each antipodal pair of sites is split across $A$ and $B$.
In the following subsections, we will show how to evaluate the algebra product $\Psi\Psi^\dagger$ and the Hilbert-Schmidt norm $\|\cdot\|$ required to calculate the purity  $\Tr[\rho^2] = ||\rho||^2$ in both the Bell and conventional basis.
We will also work through some examples and provide an informal treatment of their asymptotics; these support our claim that they exhibit the full range of area, log, and volume-law scaling.

\subsection{Bell basis method}

The symmetric tensor states in Bell basis, $\ket{\Psi^{\text{sym}}_{\nx\!,\,\ny\!,\,\nz}}$ (for brevity, we use $\ket{\Psi}$) are given by
\begin{align}
  \ket{\Psi}
  \,&{=}\,
  \frac{1}{\sqrt{N!\,n!}}
  {\sum_{\pi\in S_N}}\!\!\pi[\ket{T^B_X}^{\!\otimes\nx} \!\!{\otimes}\! \ket{T^B_Y}^{\!\otimes\ny} \!\!{\otimes}\! \ket{T^B_Z}^{\!\otimes\nz}]\text{,}
\end{align}
where $n!\,{=}\,\nx!\,\ny!\nz!$ is notational short-hand.
We use the Choi-Jamiolkowski isomorphism~\cite{Bergholm:2011} to convert this state into an operator that maps from one half-system to the other,
\begin{align}
    \Psi&=\frac{1}{\sqrt{2^N N!\,n!}}\sum_{\pi\in S_N}\pi[\sigma_X^{\otimes \nx}\otimes \sigma_Y^{\otimes \ny}\otimes \sigma_Z^{\otimes \nz}].
\end{align}
Since all the matrices in \Cref{eq:reps} are Hermitian, we have $\Psi^\dag=\Psi$ for all the states.
Our goal will be to calculate the norm $\|\rho\|^2$ of the density matrix $\rho = \Psi\Psi^\dagger$.

First, we distribute the product in the density matrix,
\begin{align}
  \rho &= \frac{1}{2^N N!\,n!}
  {\sum_{\pi,\pi'\in S_N}}\!\!
  \pi[\mediumotimes_\alpha \sigma_\alpha^{\otimes n_\alpha}]
  \pi'[\mediumotimes_\beta \sigma_\beta^{\otimes n_\beta}]\text{,}
  \label{eq:rhoB}
\end{align}
where $\alpha$ and $\beta$ go through the index set $\{X,Y,Z\}$.
We understand this formula as concerning two sets of $N$ points, called the left and right sets, which represent the tensor-product factors in \Cref{eq:rhoB}, each of which is divided into three classes $X$, $Y$ and $Z$ of size $n_X$, $n_Y$, and $n_Z$, respectively.
Each permutation in the double sum is equivalent to a labeling for one of these sets by $1$ through to $N$.
For each term, we then form a matching between left and right points by connecting those with a common label, creating a labeled matching between the two sets.
The label of a pair is the tensor-product factor in which the result of its product is placed.
We can then classify different kinds of terms using a matrix $P$ which counts the number of pairs between the different classes of the left and right sets,
\begin{align}
  P&=
  \begin{pmatrix}
    P_{XX} & P_{XY} & P_{XZ} \\
    P_{YX} & P_{YY} & P_{YZ} \\
    P_{ZX} & P_{ZY} & P_{ZZ}
  \end{pmatrix},
\end{align}
with matrix element $P_{\alpha\beta}$ the number of $\sigma_\alpha\sigma_\beta$ products in the term.
This is motivated by each term with a given $P$ matrix being equivalent, up to a reordering of the tensor factors in the result.

The collection of terms has a symmetry group -- which does not disturb the $P$ classification -- consisting of permutation actions on each of the left and right sets, restricted to leaving the three classes of points invariant, together with renumbering the labeling.
The order of the symmetry group is $(n!)^2 N!$.
Each term has a stabilizer subgroup under which it remains invariant, this is formed by the simultaneous use of the previously described left and right permutation actions to exchange equivalent strands, thereby shuffling the labeling, followed by using the renumbering action to restore the original labeling.
The order of the stabilizer subgroup is $P! = \prod_{\alpha,\beta} P_{\alpha,\beta}!$.
Therefore, using the orbit-stabilizer theorem, the summation in \Cref{eq:rhoB} can be written as,
\begin{align}
  \rho &= \frac{n!}{2^N N!}\sum_{P \in \mathcal{P}}\frac{1}{P!}
  \sum_{\pi\in S_N}
  \pi[\mediumotimes_{\alpha,\beta}(\sigma_{\alpha}\sigma_{\beta})^{\otimes P_{\alpha\beta}}],
  \label{eq:rhoB2}
\end{align}
where the $N!$ factor from the symmetry group has become the number of terms in the sum over $S_N$.

The result of the product $\sigma_{\alpha}\sigma_{\beta}$ is obtained from the multiplication table below,
\begin{center}
    \begin{tabular}{ c|ccc }
    $\times$ & $\sigma_X$ & $\sigma_Y$ & $\sigma_Z$ \\
    \hline
    $\sigma_X$ & $\mathbb{I}$ & $i\sigma_Z$ & $-i\sigma_Y$  \\
    $\sigma_Y$ & $-i\sigma_Z $ & $\mathbb{I}$ & $i\sigma_X$\\
    $\sigma_Z$ & $i\sigma_Y$ & $-i\sigma_X$ & $\mathbb{I}$ \\    
    \end{tabular}
\end{center}
for which each element is in correspondence with the matrix element of $P$ counting the number of copies of that particular pair.
Then each $P$ term in the summation in \Cref{eq:rhoB2} can be labeled by a vector $r =(r_I, r_X, r_Y, r_Z)$, where $r_{\gamma}$ denotes the number of $\sigma_\gamma$ Pauli matrices in that term.
Let $\mathcal{R}$ be the set of allowed $r$ vectors.
Note that, different $P$ matrices can yield the same $r$ vector, which forms equivalence classes we denote by $\mathcal{P}_r$.
Therefore, the summation in \Cref{eq:rhoB2} can be reassociated as,
\begin{align}
  \rho&=\frac{n!}{2^N N!}\sum_{r \in \mathcal{R}}\sum_{P \in \mathcal{P}_r}\frac{i^{\phi(P)}}{P!}
  {\sum_{\pi\in S_N}}\pi[\mediumotimes_{\gamma} \sigma_\gamma^{\otimes r_\gamma}]\text{,}
  \label{eq:rhoB3}
\end{align}
using a phase-function $\phi(P)$,
\begin{align}
  \phi(P)&= P_{XY} {+} P_{YZ} {+} P_{ZX} {-} P_{XZ} {-} P_{ZY} {-} P_{YX}\text{,}
\end{align}
which collects together the phase factors from each of the pair products.

Now, the Hilbert-Schmidt norm of $\rho$ can now be calculated readily, since contributions with different $r$ are orthogonal and the norm of each $r$ term can be calculated in the same way as the state normalization factor,
\begin{align}
  \|\rho\|^2 &=
  \frac{(n!)^2}{2^N N!}\sum_{r \in \mathcal{R}}r!\,
  \Big|\sum_{P \in \mathcal{P}_r}\frac{i^{\phi(P)}}{P!}\Big|^2\text{.}
\end{align}
where $r!\,{=}\,r_I!\,r_X!\,r_Y!\,r_Z!$.
Indeed, this was the purpose in introducing the $r$ classes to write the sum in an orthogonal and linearly independent basis.
Clearly, this summation can be evaluated in polynomial time, because the $P$ matrices are only polynomially many, which provides a method to calculate the R\'enyi entropy $S^{(2)}$ for large systems.

\subsection{Bell basis examples}

For an example let us consider the case when $n_Z = 0$ and without loss of generality take $\nx \le \ny$.
The allowed $P$ matrices and the corresponding $r$ vectors are the following; for any $a \le \nx$,
\begin{align}
 P \,{=} \begin{pmatrix}
     \nx{-}a & a & 0 \\
     a & \ny{-}a & 0\\
     0 & 0 & 0
  \end{pmatrix}\text{,}
  \quad
  r \,{=}\, (N{-}2a,0,0,2a)\text{.}
\end{align}
Thus, the number of allowed $P$ matrices is $n_X+1$.
The number of allowed $r$ vectors is also the same since each $P$ generates a distinct $r$ in this case.
$\phi(P)=0$ (since $P_{XY}=P_{YX}$) for all allowed $P$.
Thus, we obtain 
\begin{align}
    ||\rho||^2
    &=\frac{(\nx!\,\ny!)^2}{2^NN!}\sum_{a=0}^{\nx}\frac{(N-2a)!(2a)!}{((\nx{-}a)!(\ny{-}a)!(a!)^2)^2}
    \text{.}
    \label{eq:Bn1n20}
\end{align}

Let us now focus on the parameter regime where $n_X, n_Y$ are of order $N$, and hence are extensive.
The summand in \Cref{eq:Bn1n20} sharply peaks up around $a=n_X/2$.
Therefore, we will keep terms only around this value of $a$, throwing away all other terms in the summation in \Cref{eq:Bn1n20}. 
In this regime, all quantities under the factorial operation are of order $N$ and so we use Stirling's approximation on all of them in the large $N$ limit.
Thus we obtain,
\begin{align}
    ||\rho||^2
    &\,{=}
    \frac{(\nx!\,\ny!)^2}{2^N N!}\frac{e^N}{(2\pi)^3}\int_0^{\nx}\frac{\sqrt{2a(N{-}2a)}}{\,a^2(\nx{-}a)(\ny{-}a)}e^{-f(a)}da
    \text{,}
    \label{eq:Stirling}
\end{align}
where $f(a) = 2(\nx{-}a) \ln(\nx{-}a) + 2(\ny{-}a) \ln(\ny{-}a) + 2a \ln(a/2) - (N{-}2a) \ln(N{-}2a)$, and we have approximated the summation by an integration.
We will evaluate the integral in \Cref{eq:Stirling} using the saddle-point approximation.
To this end, we first note that the minimum of the function $f$ is given by 
\begin{align}
    f'(a_0)=2\ln \frac{a_0(N-2a_0)}{2(\nx-a_0)(\ny-a_0)}=0,\nonumber\\
    \implies 4a_0^2-3(\nx+\ny)a_0+2\nx\ny=0.
\end{align}

Now for the case where $\nx\approx\ny\approx N/2 + O(1)$, the minima is at $a_0=\nx/2$.
We confirm, $f''(a_0)=8/\nx>0$.
We also note that the function in the integrand in \Cref{eq:Stirling} (multiplying $e^{-f(a)}$) is almost a constant ($=16/\nx^3$) around $a=\nx/2$. This gives
\begin{align}
    ||\rho||^2&\approx\frac{(\nx!)^4}{2^NN!}\frac{16e^Ne^{-f(a_0)}}{(2\pi)^3\nx^3}\int_0^{\nx}e^{-\frac{1}{2}(a-a_0)^2f''(a_0)}da\nonumber\\
    &\approx\frac{(2\pi\nx)^2(\frac{\nx}{e})^{4\nx}}{2^N\sqrt{2\pi N}(\frac{N}{e})^N}\frac{16e^N2^{4\nx}\nx^{-2\nx}}{(2\pi)^3\nx^3}\frac{\sqrt{\pi \nx}}{2}\text{Erf}[\sqrt{\nx}]\nonumber\\
    &\approx \frac{4}{\pi N},
\end{align}
where on the last line we have used $\text{Erf}[x]=1$ for large $x$. This yields $S^{(2)}=\ln N-\ln(4/\pi)$. So, the entanglement scaling is logarithmic when $\nx=\ny$ (the result holds true for $\nx\simeq \ny$). For example, we numerically calculate $S^{(2)}$ for the state $(N\pm1,N\mp1,0)/2$ and fitting with the form $S^{(2)}=\ln N-c$ yields $c\approx0.23$ which is very close to the analytical value $\ln(4/\pi)(\approx0.24)$.

The case $\nx \,{\neq}\, \ny$ was numerically found to yield the volume law in the main text. We leave a full analytical calculation for future work.



\subsection{Conventional basis method}

Turning now to the conventional basis, we once again start by using the Choi-Jamiolkowski isomorphism~\cite{Bergholm:2011} to turn the state into an operator $\Psi$, take the density matrix $\rho = \Psi \Psi^\dagger$ and then distribute the product,
\begin{align}
  \rho
  \,{=}\,\frac{1}{2^{n_z} N!\,n!}
  \!{\sum_{\pi,\pi'\in S_N}}\!\!\!
  \pi[\mediumotimes_\alpha \sigma_\alpha^{\otimes n_\alpha}]
  \pi'[\mediumotimes_\beta \sigma_\beta^{\dagger\otimes n_\beta}]
  \text{,}
  \label{eq:rhoC}
\end{align}
where $n!\,{=}\,n_{+}!\,n_{-}!\,n_Z!$ and $\alpha$, $\beta$ now go through the index set $\{+,-,Z\}$.
Again, we classify the terms by use of a $P$ matrix,
\begin{align}
P&=
\begin{pmatrix}
  P_{++} & P_{+-} & P_{+Z} \\
  P_{-+} & P_{--} & P_{-Z} \\
  P_{Z+} & P_{Z-} & P_{ZZ}
\end{pmatrix}\text{,}
\end{align}
with matrix elements counting different types of pairs of operators in the corresponding labeled pairing.
The multiplication table for the conventional basis is given by,
\begin{center}
    \begin{tabular}{c|ccc}
    $\times$ & $(\sigma_+)^\dagger$ & $(\sigma_-)^\dagger$ & $\sigma_Z$  \\
    \hline
    $\sigma_+$ & $\mathbb{I}_\uparrow$ & 0 & $-\sigma_+$ \\
    $\sigma_-$ & $0$ & $\mathbb{I}_\downarrow$ & $\sigma_-$ \\
    $\sigma_Z$ & $-\sigma_-$ & $\sigma_+$ & $\mathbb{I}$
    \end{tabular}
    \text{,}
\end{center}
where $\mathbb{I}_{\uparrow}=(\mathbb{I}+\sigma_Z)/2$ and $\mathbb{I}_{\downarrow}=(\mathbb{I}-\sigma_Z)/2$ are projectors into the $Z$-basis.
Note that, any term with nonzero $P_{+-}$ or $P_{-+}$ will be identically zero, which further reduces the allowed $P$ matrices, in addition to the previous constraints on the row and column sums.
Furthermore, we can ignore the minus signs in this table because the constraints on $P$ force $P_{+Z}=P_{Z+}$.

Unlike in the Bell basis, the different operators seen in this table do not immediately generate an orthonormal basis.
Instead we choose an orthogonal basis $\{\sigma_+,\sigma_-,\mathbb{I}_\uparrow,\mathbb{I}_\downarrow\}$ for the operator products, and then expand each term in that basis using $\mathbb{I} = \mathbb{I}_\uparrow + \mathbb{I}_\downarrow$.
This choice is not unique and alternatives, such as expanding in the $\{\mathbb{I},\sigma_Z\}$ basis, may be advantageous depending on the state.
The terms of that expansion can be labeled by vectors $r = (r_{+}, r_{-}, r_\uparrow, r_\downarrow)$ for the number of factors of each basis operator in the resulting tensor-product.
Unlike in the Bell basis case, each $P$ class of terms can appear in multiple different $r$ classes after this additional expansion.
We now reassociate the density matrix sum by $r$,
\begin{align}
  \rho &\,{=}\, 
  \frac{n!}{2^{n_Z} N!}\sum_{r \in \mathcal{R}}\sum_{P \in \mathcal{P}_r}
  \frac{1}{P!}
  \binom{P_{ZZ}}{r_\uparrow {-} P_{++}}
  \smashoperator{\sum_{\pi \in S_N}}
  \pi[\mediumotimes_{\gamma}\sigma_\gamma^{r_\gamma}]
  \label{eq:rhoC2}
\end{align}
where $\gamma \in (+,-,\uparrow,\downarrow)$ and the binomial factor comes from expanding $\mathbb{I}$ as discussed.

We can now find an expression for the norm by using the orbit-stabilizer theorem again as in the state normalization calculation,
\begin{align}
  \|\rho\|^2 &= \frac{(n!)^2}{2^{2n_Z} N!} \sum_{r \in \mathcal{R}} r!\,\Big|\sum_{P\in \mathcal{P}_r}\!\frac{1}{P!}\,\binom{P_{ZZ}}{r_\uparrow - P_{++}} \Big|^2
\end{align}
where $r!\,{=}\,r_+!\,r_-!\,r_\uparrow!\,r_\downarrow!$.
At this point, we are ready to evaluate this expression and calculate the R\'enyi entropy, as it can clearly be done in polynomial time.

\subsection{Conventional basis examples}
Let us take a simple example where $\nz = 0$.
The only allowed $P$ matrix in this case is 
\begin{align}
  P = \begin{pmatrix}
    n_{+} & 0 & 0\\
    0 & n_{-} & 0 \\
    0 & 0 & 0 
  \end{pmatrix}\text{,}
\end{align}
which corresponds to $r = (0, 0, n_{+}, n_{-})$.
The reduced density matrix is given by,
\begin{align}
  \rho&=\frac{1}{N!}\sum_{\pi \in S_N}\pi[\mathbb{I}_\uparrow^{\otimes n_{+}}\otimes\mathbb{I}_\downarrow^{\otimes n_{-}}]\text{.}
\end{align}
Thus, we obtain 
\begin{align}
  \|\rho\|^2=\frac{n_{+}!n_{-}!}{N!}\text{,}
\end{align}
where the $n!$ factor comes from the order of the stabilizer group.
This yields the R\'enyi-2 entropy $S^{(2)} = \ln N! - \ln n_{+}! - \ln n_{-}!$.
For small $n_{-} = O(1)$,
we obtain $S^{(2)} = n_{-} \ln N - \ln n_{-}!$ and the state is logarithmically entangled.
But when
both $n_{+}$ and $n_{-}$ scales with $N$ extensively, the entanglement follows a volume law.
For example, if both $n_{+}$ and $n_{-}$ are within $O(1)$ of $N/2$ then $S^{(2)} = N \log 2 - O(1/N)$, hence these states are nearly maximally entangled.
In general, the coefficient of the volume scaling is the binary entropy for the mixture between $n_{+}$ and $n_{-}$.

Let us take another example by considering those states with $n_{-}=0$, which can be indexed by a choice of $\nz(\le n_+)$.
In this case, we have $(\nz+1)$ allowed $P$ matrices which are (along with the corresponding $r$ vectors) are given by,
\begin{align}
  P &\,{=} \begin{pmatrix}
    n_{+}{-}a & 0 & a \\
    0 & 0 & 0 \\
    a & 0 & \nz{-}a
  \end{pmatrix}\text{,}&
  r &\,{=} (a,a,N{-}2a{-}b,b)\text{,}
\end{align}
where $b$ indexes the expansion of the resulting operator from $P$ into the orthogonal basis of $\{\sigma_{+},\sigma_{-},\mathbb{I}_\uparrow,\mathbb{I}_\downarrow\}$.

The reduced density matrix is now given by,
\begin{align}
\rho &= \frac{n!}{N!2^\nz} \sum_{a=0}^{\nz} \sum_{\pi \in S_N} \!\!\pi [\sigma_+^{\otimes a} \otimes \sigma_-^{\otimes a} \otimes
\mathbb{I}_\downarrow^{\otimes (n_+ -a)}\otimes
\mathbb{I}^{\otimes (\nz-a)}]
\text{.}
\end{align}
The calculation of $\|\rho\|^2$ is involved due to the presence of cross terms. We find,
\begin{align}
    \|\rho\|^2&=2^{-2\nz}\binom{N}{\nz}^{-2}\sum_{a=0}^{\nz}\frac{N!}{(N-\nz-a)!(\nz-a)!(a!)^2}\nonumber\\
    &\times \sum_{i=0}^{\nz-a}\binom{\nz-a}{i}\binom{N-\nz-a}{i}2^{\nz-a-i}\nonumber\\
    &=N!\binom{N}{\nz}^{-2}\sum_{k=0}^{\nz}\frac{2^{-k}\Gamma(\nz-k+\frac{1}{2})}{\sqrt{\pi}k!((\nz-k)!)^3(N-2\nz+k)!},
    \label{eq:Tc1}
\end{align}
where on the second line, we have used the following change of variables: $k=\nz-i-a$. We are interested in the scaling behavior of $S^{(2)}$ in two separate parameter regime which we analyze below. 

First, we focus on the asymptotic behavior (i.e., $N \to \infty$) of $S^{(2)}$ with constant $\nz$ ($\sim O(1)$). Let us first simplify \Cref{eq:Tc1} by using Stirling's approximation and keeping only the leading order terms. Thus, we obtain
\begin{align}
  \|\rho\|^2&\approx(\nz!)^2\sum_{k=0}^{\nz}\frac{2^{-k}\Gamma(\nz-k+\frac{1}{2})N^{-k}}{\sqrt{\pi}k!((\nz-k)!)^3}.
\end{align}
The leading behavior comes from the first term ($k=0$) which gives
\begin{align}
  \|\rho\|^2&=\frac{\Gamma[\nz+1/2]}{\sqrt{\pi}\nz!}+O(1/N).
\end{align}
Hence the entanglement entropy is
\begin{align}
   \lim_{N\to\infty}S^{(2)} =\ln\left(\frac{\sqrt{\pi}\nz!}{\Gamma[\nz+1/2]}\right),
\end{align}
and follows the area law in the asymptotic limit.

Second, we analyze the behavior of $S^{(2)}$ when $\nz/N(=x)$ is a nonzero fraction in the $N \to \infty$ limit. In this case, the summand in \Cref{eq:Tc1} peaks up around a value of $k$ which is a fraction (proportional to x) of $\nz$. Hence, all the quantities inside the factorial symbol can be expanded using the Stirling's approximation in the large $N$ limit. Thus we get,
\begin{align}
    ||\rho||^2&=\frac{N!}{(2\pi)^2}\binom{N}{\nz}^{-2}\smashoperator{\int_0^{\nz}}\frac{e^{-f(k)}dk}{\sqrt{\pi k (\nz-k)^3(N-2\nz+k)}},
    \label{eq:Tcintegral}
\end{align}
where $f(k)=k\ln(2k)+2(\nz-k)\ln(\nz-k)+(N-2\nz+k)\ln(N-2\nz+k)-N$. We find the function $f(k)$ shows a minima at $k_0=N[x-1+\sqrt{(x-1)^2+x^2}]$ which is confirmed by $f''(k_0)>0, \ \forall x\in[0,1]$. The function in the denominator of the integrand (say, $w(k)$) in \Cref{eq:Tcintegral} is almost a constant around $k_0$ (where the numerator peaks up) and hence can be taken out of the integral. Thus, we get
\begin{align}
   ||\rho||^2&\approx\frac{N!}{(2\pi)^2}\binom{N}{\nz}^{-2} \frac{e^{-f(k_0)}}{w(k_0)}\int_0^{\nz}e^{-\frac{1}{2}(k-k_0)^2f''(k_0)}dk,
\end{align}
where, for $x=1/2$, we get
\begin{align}
k_0&=N(\sqrt{2}-1)/2,\nonumber\\
f(k_0)&=-N(2+2\sqrt{2}\ln(1+\sqrt{2})+\sqrt{2} \ln(3-2\sqrt{2})\nonumber\\
&\quad+\ln(6+4\sqrt{2})-2\ln N)/2 ,\nonumber\\
f''(k_0)&=(8+6\sqrt{2})/N ,\nonumber\\
w(k_0)&=\left(\frac{\pi N^5}{464+328\sqrt{2}}\right)^{1/2}.
\end{align}
which yields
\begin{align}
    S^{(2)}&\approx N\ln\left(\frac{2(2-\sqrt{2})}{(3-2\sqrt{2})^{\frac{1}{\sqrt{2}}}(1+\sqrt{2})^{\sqrt{2}}}\right)+\frac{1}{2}\ln N-0.06.
    \label{eq:Tc2}
\end{align}

We numerically calculate $S^{(2)}$ for the state $(N-1,0,N+1)/2$ and fitting with the form $S^{(2)}=c_1N+c_2\ln N+c_3$ yields $(c_1,c_2)=(0.15837, 0.492)$ which are very close to the analytical values (0.15835, 0.5) from \Cref{eq:Tc2}.

\section{Correlation functions}
\label{app:Cl}

Here we discuss the behavior of two-point correlation function $\mathcal{C}[l]=\langle {\bf S}_i\cdot{\bf S}_{i+l}\rangle$. We first prove a simple yet important relation between $\mathcal{C}[l]$ and $\langle S^2_{\text{tot}}\rangle$ of a symmetric tensor state. Let us start with
\begin{align}
    S^2_{\text{tot}}=\sum_{i,j=1}^{2N}{\bf S}_i\cdot{\bf S}_j&=\frac{3N}{2}+2\sum_{i<j}{\bf S}_i\cdot{\bf S}_j,
\end{align}
where 
\begin{align}
   \sum_{i<j}{\bf S}_i\cdot{\bf S}_j&=\sum_{i=1}^N {\bf S}_i\cdot{\bf S}_{i+N}+2\sum_{i=1}^N\sum_{l=1}^{N-1}{\bf S}_i\cdot{\bf S}_{i+l} .
\end{align}

We know $\mathcal{C}[l]$ is independent of $l$ and the value of the antipodal correlations ($\mathcal{C}[N]$) are 1/4. This gives us
\begin{align}
   \langle S^2_{\text{tot}}\rangle&=\frac{3N}{2}+\frac{N}{2}+4N(N-1)\mathcal{C}[l]\nonumber\\
   \therefore \mathcal{C}[l]&=\frac{\langle S^2_{\text{tot}}\rangle}{4N(N-1)}-\frac{1}{2(N-1)}.
\end{align}
So, $\mathcal{C}[l]$ and $\langle S^2_{\text{tot}}\rangle$ follow a linear relationship and one can be obtained from the other. In this appendix, we explicitly calculate the local correlation functions ($\mathcal{C}[1]$) for the symmetric tensor states in both the Bell and conventional bases. 

\subsection{Bell basis}
We first note that,
\begin{align}
    \bra{T^B_\alpha T^B_\beta}{\bf S}_i \cdot {\bf S}_{i+1}\ket{T^B_\beta T^B_\alpha} &=
    \begin{dcases}
      0 &: \quad \alpha = \beta\text{,}\\
      \frac{1}{4}&: \quad \alpha \neq \beta{,}\\
    \end{dcases}
\end{align}
where $i$ is the site index and $\alpha,\beta \in \{X,Y,Z\}$ are indices into the Bell basis.
Therefore, for the state $(\nx,\ny,\nz)$, we obtain
\begin{align}
  \mathcal{C}[1]&=
  \frac{(N{-}2)!}{N!}
  \sum_{\alpha,\beta}
  \binom{n_\alpha}{1}\binom{n_\beta}{1}
  \bra{T^B_\alpha T^B_\beta}{\bf S}_i \cdot {\bf S}_{i+1}\ket{T^B_\alpha T^B_\beta}\nonumber\\
  &=\frac{\nx\ny+\ny\nz+\nz\nx}{2N(N-1)}
  \text{.}
  \label{eq:Bn1n2n3}
\end{align}
Therefore, to give a few examples, for the state ($2N/3,N/3,0$), $\mathcal{C}[1]=N/(9N-9)$ while for the state ($N/3,N/3,N/3$), $\mathcal{C}[1]=N/(6N-6)$. Thus, these states have non-thermal expectation values (for the local observable $\mathcal{C}[1]$) even in the thermodynamic limit.

\subsection{Conventional basis}

First, we note that
\begin{align}
  \bra{T^C_ZT^C_Z}{\bf S}_i \cdot {\bf S}_{i+1}\ket{T^C_ZT^C_Z}&=0,\nonumber\\
  \bra{T^C_\alpha T^C_\alpha}{\bf S}_i \cdot {\bf S}_{i+1}\ket{T^C_\alpha T^C_\alpha}&=+\frac{1}{4}\text{,}\nonumber\\
  \bra{T^C_\alpha T^C_Z}{\bf S}_i \cdot {\bf S}_{i+1}\ket{T^C_ZT^C_\alpha}&=+\frac{1}{4}\text{,} \nonumber\\
  \bra{T^C_\alpha T^C_\beta}{\bf S}_i \cdot {\bf S}_{i+1}\ket{T^C_\beta T^C_\alpha}&=-\frac{1}{4} \quad \text{(for $\alpha \neq \beta$),}
\end{align}
where $\alpha,\beta \in \{{+},{-}\}$ are indices into the conventional basis.
So, for the general state $(n_{+},n_{-},\nz)$ we obtain
\begin{align}
    \mathcal{C}[1]&=
    \frac{(N{-}2)!}{N!}
    \sum_{\alpha,\beta}\mathcal{M}_{\alpha\beta}\bra{T^C_\alpha T^C_\beta}{\bf S}_i \cdot {\bf S}_{i+1}\ket{T^C_\alpha T^C_\beta}\nonumber\\
    &=\frac{(n_{+}-n_{-})^2+(n_{+}+n_{-})(2\nz-1)}{4N(N-1)},
    \label{eq:Cn1n2n3}
\end{align}
where
\begin{align}
    \mathcal{M}_{\alpha\beta}&= \begin{dcases}
      \binom{n_\alpha}{1}\binom{n_\beta}{1} &:\quad \alpha \neq \beta\text{,} \\
     2 \binom{n_\alpha}{2} &:\quad \alpha = \beta\text{.}
    \end{dcases}
\end{align}
Note that, the result in \Cref{eq:Cn1n2n3} is invariant under $n_{+} \leftrightarrow n_{-}$.
Now, for the states $(N-c,c,0)$, we obtain
\begin{align}
   \mathcal{C}[1]&=\frac{1}{4}\Big(1-\frac{4c}{N}\Big)\Big(1-\frac{c-1}{N-1}\Big).
\end{align}
Therefore, this state has a non-thermal local expectation value in the thermodynamic limit. But for states $(N\pm1, N\mp1, 0)/2$, we obtain $\mathcal{C}[1]=-1/(4N)$ which is thermal in the thermodynamic limit.

\section{Von Neumann entanglement entropy}
\label{app:vN}

\begin{figure}[t]
    \centering
    \includegraphics[width=\linewidth]{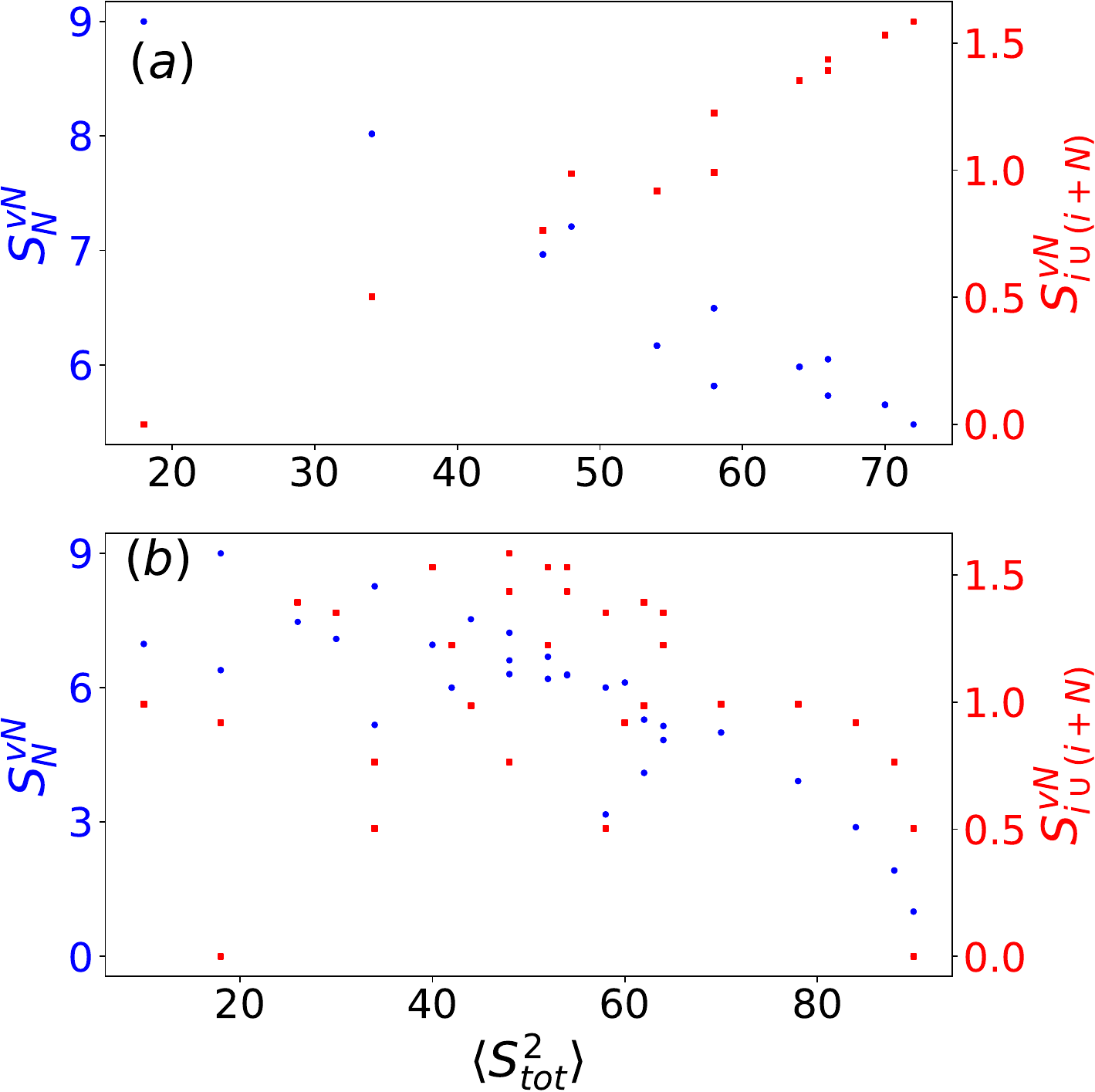}
    \caption{Behavior of von-Neumann entanglement entropy for different cuts for (a) Bell and (b) Conventional basis. The system size is $2N=18$.}
    \label{fig:EE}
\end{figure}

In this appendix, we discuss the behavior of the von Neumann entanglement entropy: $S^{\text{vN}}_A\,{=}\,-\Tr[\rho\ln \rho]$ where $\rho$ is the reduced density matrix of subsystem $A$. To this end, we consider two different bipartition schemes, the usual contiguous one such as half-chain entanglement entropy $S^{\text{vN}}_N$, and the entanglement of two antipodal sites with the rest of the system ($S^{\text{vN}}_{i\cup[i+N]}$). The value of the former is $N\ln 2$ and hence extensive but the latter is zero for the entangled-antipodal-pair root states. The former (latter) is found to exhibit a decreasing (increasing) trend with $\langle S^2_{\text{tot}}\rangle$, particularly in Bell basis [\Cref{fig:EE}(a)]. While the increase of $S^{\text{vN}}_{i\cup[i+N]}$ from zero destroys the product structure of the states and drifts them towards more typical/generic behavior, the simultaneous decrease of $S^{\text{vN}}_N$ gradually makes them more and more atypical. Such pattern is not prominent in the conventional basis [\Cref{fig:EE}(b)] but here also the states with maximum $S^{\text{vN}}_{i\cup[i+N]}$ have significantly reduced $S^{\text{vN}}_{N}$.

\begin{figure}
   \centering
   \includegraphics[width=\linewidth]{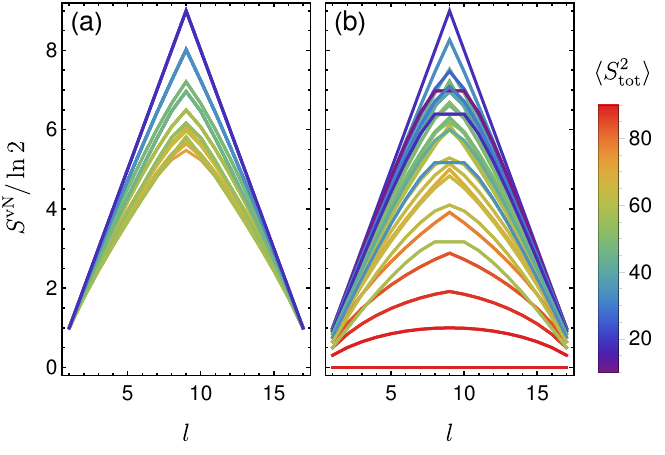}
   \caption{Von Neumann entanglement entropy ($S^{\text{vN}}$) as a function of subsystem size ($l$) for the symmetric tensor states in (a) Bell and (b) conventional basis. The system size is $2N=18$.}
   \label{fig:entscaling}
\end{figure}

The scaling of $S^{\text{vN}}$ vs $l$ exhibits different behavior in the two bases of triplets. In the Bell basis, the decrease in entanglement with $\langle S^2_{\text{tot}}\rangle$ appears to slow down and saturate at a non-zero value (see \Cref{fig:entscaling}). In the conventional basis, two ferromagnetic root states ($\ket{\Psi(T^C_+)}, \ket{\Psi(T^C_-)}$) have zero entanglement and the $\ket{\Psi(T^C_Z)}$ is maximally entangled. The $S^{\text{vN}}$ of all other states in this basis ranges almost uniformly between these two extreme values (see \Cref{fig:entscaling}). Therefore, the states are more scarred in the conventional basis compared to the Bell basis. We find, that the entanglement minimization~\cite{Reuvers2018Jun} within the symmetric tensor manifold in the Bell basis extracts the maximal spin components from each basis state and combines them together to create the ferromagnetic vacua. 
Entanglement entropy is also found to be exactly the same for different states with the same $\langle S^2_{\text{tot}} \rangle$, with a few exceptions. Unlike the correlation functions, some states with the same $\langle S^2_{\text{tot}}\rangle$ have different entropy due to the presence of different numbers of entangled antipodal pair products (i.e.\@ different $\mathcal{N}_c$). Such states are found to appear for $N\geq 9$.

\section{Stability against perturbations}
\label{app:stability}
In this appendix, we study the stability of symmetric tensor scars (STS) against different perturbations\footnote{Here we use the word ``perturbations" to refer to some physical terms (hermitian operators). We do not restrict ourselves to some small values of such terms. In fact, nothing is small in the middle of the spectrum since the energy gaps are exponentially small in $L$.}. In Bell basis, each triplet root state is annihilated by one of the $S^{\alpha}_{\text{tot}}$ operators: $S^y_{\text{tot}} \ket{\Psi(T^B_X)} = 0$, $S^x_{\text{tot}} \ket{\Psi(T^B_Y)} = 0$, $S^z_{\text{tot}} \ket{\Psi(T^B_Z)} = 0$, hence they are perfectly stable (i.e. remain zero energy eigenstates) against arbitrarily large global field in the corresponding direction. Moreover, these states are perfectly stable even when the staggered n.n.\@ exchange interaction is anisotropic in all three directions. Many STS are also found to be zero energy eigenstates of staggered field ($P_1=h_z\sum_i(-1)^iS^z_i$) and easy-axis anisotropy  ($P_2=-J_z\sum_i(-1)^iS^z_iS^z_{i+r}$) as charted out in \Cref{tab:stability}.
In the conventional basis, all STS are eigenstates of $S^z_{\text{tot}}$.

\begin{table}[ht!]
  {\setlength{\tabcolsep}{6pt}
  \begin{tabular}{llll}
  \hline \hline
    & \multicolumn{3}{c}{Perturbation to \Cref{eq:SH}} \\
    \cline{2-4}
    $N$   &  0 & $P_1$ & $P_2$ \\
    \hline
      3 & 11 & 4 & 7  \\
      5 & 22 & 6 & 8  \\
      7  & 37 & 8 & 10  \\
      9 & 56 & 10 & 12 \\
      \hline \hline
  \end{tabular}}
  \caption{
    The number of symmetric tensor states, of the Bell basis, perfectly stable against (i.e remain zero energy eigenstates) different perturbations $P_1$ and $P_2$ to the Hamiltonian from \Cref{eq:SH} in the main text. 
  }
  \label{tab:stability}
\end{table}

We now study the effect of the perturbations ($P_1$ and $P_2$) on the STSs which are not exact eigenstates of these perturbations. Any generic perturbation is expected to split the huge degeneracy of the zero energy manifold. But the later is protected by a combination of symmetries and as long as they are intact, the zero energy manifold exist (though the size of it may vary). Therefore, let us first analyze the symmetry properties of our chosen perturbations. Both $P_1$ and $P_2$ breaks continuous spin rotation SU(2) symmetry but invariant under the U(1) symmetry corresponding to the conservation of magnetization. Moreover, we have $\{P_{1,2},\tau\}=0$, $[P_{1,2},\tau^2]=0$, $[P_1,I]\neq 0$ but $[P_2,I]=0$ where $\tau$ is the translation by one lattice spacing and $I$ is the space inversion operator. This is why the zero energy manifold persists under the addition of these perturbations, although the size reduces (more for $P_1$ because of the absence of the inversion symmetry). We find (see \Cref{tab:stability}) that for $L=14$ ($N=7$) total 29 (27) STS are destabilized under the addition of $P_1$($P_2$) (i.e they are no longer eigenstates of the new Hamiltonian $H'=H+P_1(P_2)$). Interestingly, the energy of these states w.r.t the new Hamiltonians remain zero for both kind of perturbations. This can be seen easily using the symmetries (all STS are zero-momentum states) or perturbative arguments~\cite{Lin_perturb}.

\begin{figure}
    \includegraphics[width=\linewidth]{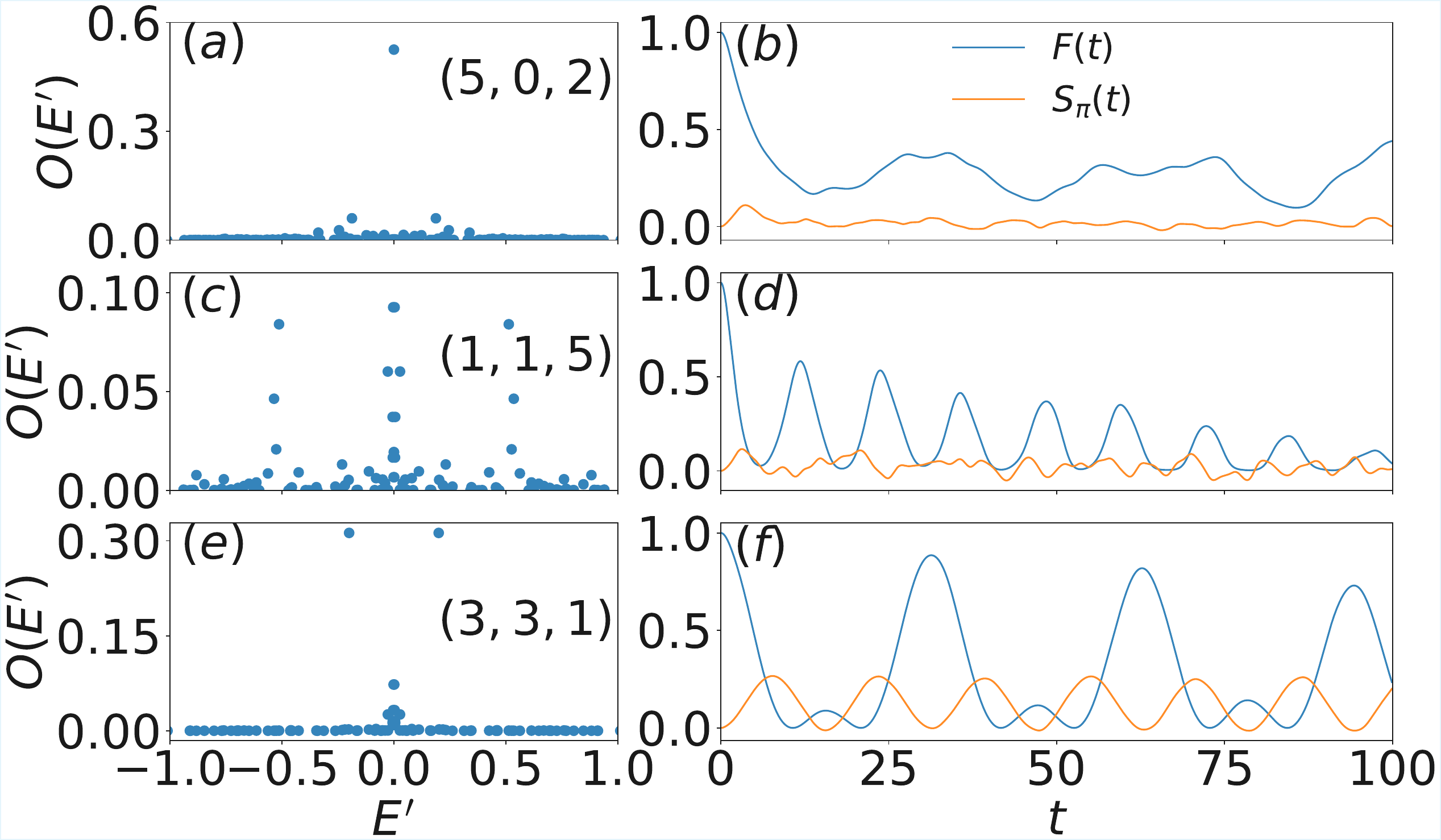}
    \caption{Stability of STS under $P_1$: Behavior of overlaps $O(E')$ and the dynamics of Fidelity ($F(t)$) and structure factor ($S_{\pi}(t)$) from the initial states $(5,0,2)$ (in panel (a),(b)), $(1,1,5)$ (in panel (c), (d)), $(3,3,1)$ (in panel (e), (f)). $L=2N=14$, $h_z=0.2$.}
    \label{fig:P1}
\end{figure}

\begin{figure}
    \includegraphics[width=\linewidth]{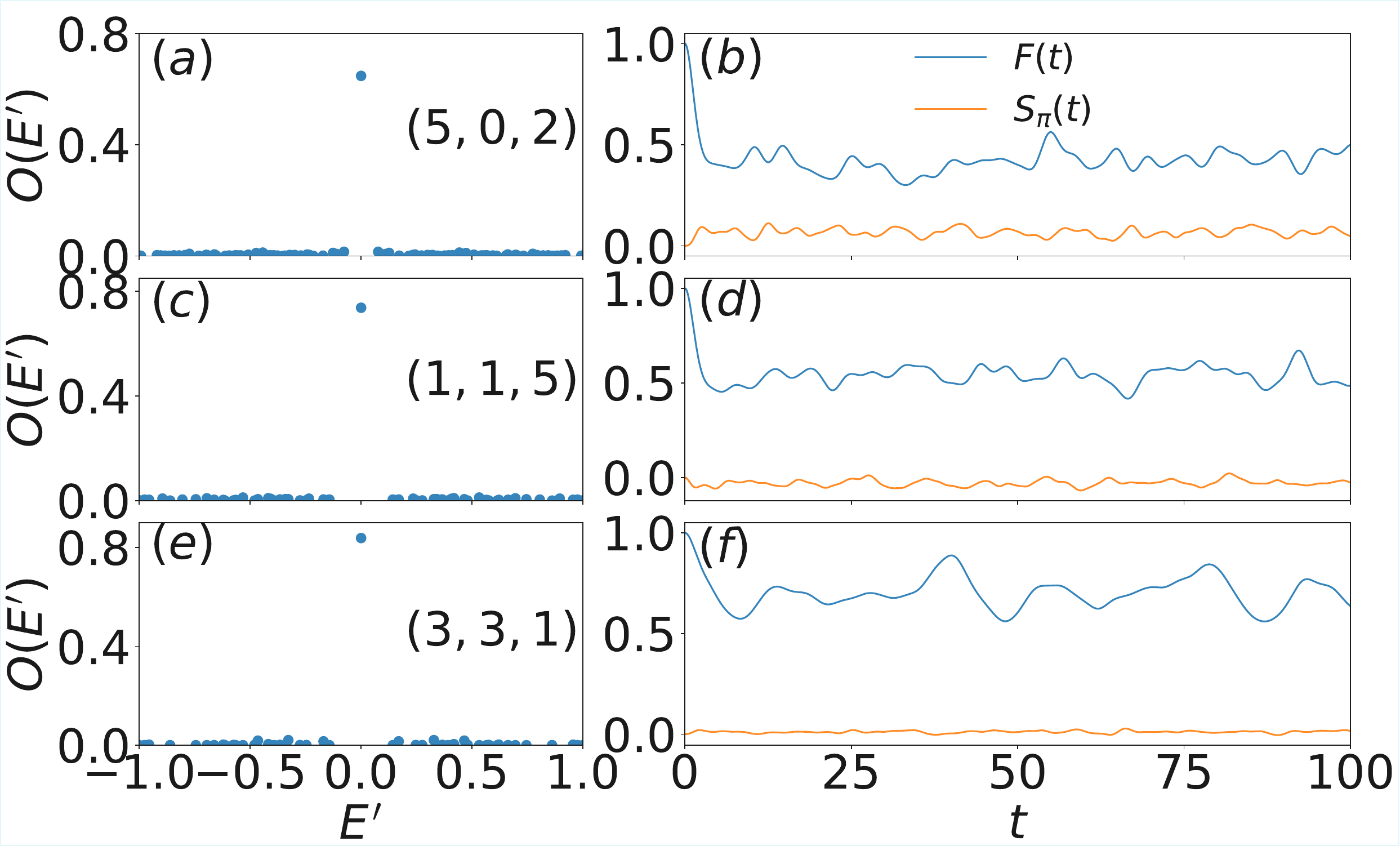}
    \caption{Stability of STS under $P_2$: Behavior of overlaps $O(E')$ and the dynamics of Fidelity ($F(t)$) and structure factor ($S_{\pi}(t)$) from the initial states $(5,0,2)$ (in panel (a),(b)), $(1,1,5)$ (in panel (c), (d)), $(3,3,1)$ (in panel (e), (f)). $L=2N=14$, $J_z=0.5$.}
    \label{fig:P2}
\end{figure}

 We now study the behavior of the overlap spectrum : $O(E')=|\bra{\Psi}E'\rangle|^2$ (where $H'\ket{E'}=E'\ket{E'}$ and $\ket{\Psi}$ is an STS in Bell basis) and the time evolution of the corresponding STS under the perturbed Hamiltonian $H'$. For $P_1$ (see Fig.~\ref{fig:P1}), we find that there are some states (like $(5,0,2)$) for which $O(E')$ is still spiked at $E'=0$. Consequently, we see slow dynamics of fidelity and other appropriate observables. In contrast, there are some states (such as $(1,1,5)$, $(3,3,1)$) which hybridize strongly with nonzero energy ($E'\neq0$) eigenstates. Since $O(E')$ is symmetric around $E'=0$ (signifying bifurcations), this leads to oscillatory dynamics when the system is unitarily evolved under $H+P_1$ starting from these STS. The oscillation from the state $(3,3,1)$ is found to be more persistent compared to $(1,1,5)$.
 We look at the time evolution of the structure factor at momentum $\pi$,
\begin{equation}
    S_{\pi}(t)=\sum_{\ell=1}^{N-1}(-1)^\ell\mathcal{C}[l,t].
\end{equation}
In consistence with the fidelity dynamics, $S_{\pi}(t)$ shows almost no change for the state $(5,0,2)$ and little change for the state $(1,1,5)$. Interestingly, it exhibits a near undamped oscillation for the state $(3,3,1)$. This means the state undergoes a coherent back-and-forth movement between a ferro- and antiferro-magnetically ordered state. We leave a detailed investigation of this behavior as an interesting future problem. Under $P_2$ (see Fig.~\ref{fig:P2}), the same STSs remain confined mostly in the new zero-energy manifold with little mixing to the other part of the spectrum (unlike $P_1$, we do not see significant bifurcations). All the STSs, hence exhibits slow dynamics under $H+P_2$. In conclusion, many STSs are quite stable against certain symmetry-breaking perturbations even when not being eigenstates of them.
\clearpage
\end{document}